\documentclass{sig-alternate-05-2015}

\usepackage{subfigure}
\usepackage{xcolor}
\usepackage{xspace}
\usepackage{courier}
\usepackage{url}
\usepackage{soul}
\usepackage{paralist}

\newif\ifdraft
\draftfalse
\ifdraft
 \newcommand{\atnote}[1]{ {\textcolor{blue} { ***AT: #1 }}}
 \newcommand{\jhanote}[1]{ \textcolor{red}  {***SJ:#1}\xspace}
\else
 \newcommand{\atnote}[1]{}
 \newcommand{\jhanote}[1]{}
\fi

\begin{document}

\title{RepEx: A Flexible Framework for Scalable Replica Exchange Molecular Dynamics Simulations}

\numberofauthors{6}

\author{
\alignauthor{Antons Treikalis}\\
       \affaddr{\small{Department of Electrical and Computer Engineering, Rutgers University}}\\
       \affaddr{\small{Piscataway, New Jersey 08854}}\\
       \affaddr{\small{antons.treikalis@rutgers.edu}}
\alignauthor{Andre Merzky}\\
       \affaddr{\small{Department of Electrical and Computer Engineering, Rutgers University}}\\
       \affaddr{\small{Piscataway, NJ 08854}}\\
       \affaddr{\small{andremerzky@gmail.com}}
\alignauthor{Haoyuan Chen}\\
       \affaddr{\small{Department of Chemistry and Chemical Biology, Rutgers University}}\\
       \affaddr{\small{Piscataway, New Jersey 08854}}\\
       \affaddr{\small{haoyuan.chen@rutgers.edu}}
\and
\alignauthor{Tai-Sung Lee}\\
       \affaddr{\small{Department of Chemistry and Chemical Biology, Rutgers University}}\\ 
       \affaddr{\small{Piscataway, New Jersey 08854}}\\
       \affaddr{\small{taisung@rutgers.edu}}
\alignauthor{Darrin M. York}\\       
       \affaddr{\small{Department of Chemistry and Chemical Biology, Rutgers University}}\\ 
       \affaddr{\small{Piscataway, New Jersey 08854}}\\
       \affaddr{\small{Darrin.York@rutgers.edu}}
\alignauthor{Shantenu Jha}\\
       \affaddr{\small{Department of Electrical and Computer Engineering, Rutgers University}}\\
       \affaddr{\small{Piscataway, NJ 08854}}\\
       \affaddr{\small{shantenu.jha@rutgers.edu}}
}

\date{18 January 2016}

\maketitle
\begin{abstract}

Replica Exchange (RE) simulations have emerged as an important algorithmic tool for the molecular sciences. RE simulations involve the concurrent execution of independent simulations which infrequently interact and exchange information. The next set of simulation parameters are based upon the outcome of the exchanges.

Typically RE functionality is integrated into the molecular simulation software package. A primary motivation of the tight integration of RE functionality with simulation codes has been performance. This is limiting at multiple levels. First, advances in the RE methodology are tied to the molecular simulation code. Consequently these advances remain confined to the molecular simulation code for which they were developed.  Second, it is difficult to extend or experiment with novel RE algorithms, since expertise in the molecular simulation code is typically required. 

In this paper, we propose the RepEx framework which address these aforementioned shortcomings of existing approaches, while striking the balance between flexibility (any RE scheme) and scalability (tens of thousands of replicas) over a diverse range of platforms.  RepEx is designed to use a pilot-job based runtime system and support diverse RE Patterns and Execution Modes.  RE Patterns are concerned with synchronization mechanisms in RE simulation, and Execution Modes with spatial and temporal mapping of workload to the CPU cores.  We discuss how the design and implementation yield the following primary contributions of the RepEx framework: (i) its ability to support different RE schemes independent of molecular simulation codes, (ii) provide the ability to execute different exchange schemes and replica counts independent of the specific availability of resources, (iii) provide a runtime system that has first-class support for task-level parallelism, and (iv) required scalability along multiple dimensions.

\end{abstract}

\section{Introduction} \label{introduction}

The Replica Exchange (RE) class of methods~\cite{swendsen1986replica} is a popular technique to enhance sampling in molecular simulations.  Although RE methods were introduced for Monte Carlo methods, their use with Molecular Dynamics (MD) has grown rapidly. There are several hundred publications every year using some variant of Replica Exchange Molecular Dynamics (REMD) in a range of scientific disciplines including chemistry, physics, biology and materials science.

REMD simulation consists of two phases: one phase is comprised of MD simulations of N different replicas of the original system, where each replica has different thermodynamic configuration.  The other phase involves exchanges of the thermodynamic configurations between replicas using Metropolis-like acceptance criterion.  Initially REMD~\cite{sugita1999replica} was used to perform exchanges of temperatures, but has since been extended to perform Hamiltonian Exchange ~\cite{fukunishi2002hamiltonian}, pH Exchange ~\cite{meng2010constant} and other exchange types.
 
Reinforcing the importance of RE, many community MD engines~\cite{Amber, namd2005, gromacs1995}, have evolved to support internal RE implementations. These solutions often demonstrate respectable performance but share a number of important limitations, many of which stem from the tight integration of RE method with the MD simulation engine.  For example, tight integration results in a significant duplication of effort between ``competing'' MD engines, or what is worse, RE algorithmic innovation that are localized to specific MD engines, as opposed to propagating across community codes.  MD engines~\cite{Amber, namd2005, gromacs1995} are highly optimized and specialized codes, often requiring hundreds of person years of development. Domain scientists are typically unprepared for the full complexity of these MD engines, however they are the ones most capable of algorithmic and methodological innovation.  Furthermore, most existing REMD implementations are not generalized: the number of exchange parameters, order and number of dimensions is hard-coded and covers a narrow spectrum of REMD simulations. Thus, the tight integration introduces a barrier to methodological advances and extensibility.

REMD simulations are characterized by multiple replicas. Depending upon the physical problem under investigation, the computational requirements of a single replica can vary from optimally running on a single core to several hundred nodes.  Additionally, depending upon number of replicas and RE scheme employed -- synchronous or asynchronous, single or multiple dimension, the number of ``active'' cores required can vary. Any REMD simulation should be expressed independent (agnostic) of the resource availability and specific resource management details, leaving the mapping of replicas to resources to a runtime system. Furthermore, any REMD framework should be capable of gracefully handling replica failures, which means that in the presence of failures, the entire simulation need not be stopped or restarted. These and other reasons reiterate the importance of an effective runtime system as simulation size and time-scales increases.

In this paper we present RepEx framework~\cite{repex-code} -- a user-level software framework with a runtime system carefully designed to support the requirements of scalable REMD simulations over multiple dimensions.  Conceptually, RepEx is designed to decouple the implementation of RE algorithm from MD simulation engine. The design of RepEx facilitates implementation of new RE algorithms with a wide range of MD engines.  The implementation of RepEx supports up to three dimensional REMD simulations with arbitrary ordering of available exchange types. 
Both \textbf{synchronous} and \textbf{asynchronous} RE simulations are supported by RepEx.

RepEx also decouples the resource management from the REMD algorithm and MD simulation engine specific details. RepEx relies on RADICAL-Pilot (RP)~\cite{review_radicalpilot_2015} as a runtime system to perform resource allocation, task scheduling and data movement. Another distinctive feature of RepEx, partly arising from the use of RADICAL-Pilot, is fault tolerance: RepEx can either continue a simulation in case of replica failure or can relaunch a failed replica. 

The functionality and flexibility come at a performance price, especially when compared to highly-customized approaches. We carefully characterize the performance ``penalty'' and argue that it is an acceptable trade-off given the functionality and flexibility enhancements proffered by RepEx. In fact, as we will chronicle, the diversity in algorithms, exchange parameters and dimensionality at adequate scales is unprecedented.

This paper is organized as follows: in Section~\ref{remd.landscape} we discuss the importance of asynchronous RE and outline existing frameworks for REMD simulations.  In Section~\ref{repex.framework} we define the requirements of REMD simulations, provide an overview of the design of RepEx (subsection ~\ref{design}) and its implementation (subsection~\ref{implementation}). We introduce two concepts -- \textbf{Replica Exchange Pattern} and \textbf{Execution Mode}. 

In Section~\ref{validation} we validate the implementation of RepEx and in Section~\ref{experiments} we present experiments conducted to demonstrate features and capabilities of our framework. We characterize the performance of RepEx for 1D REMD and REMD simulations in multiple dimensions (M-REMD). We present results using two of the most popular MD engines -- Amber and NAMD, and demonstrate capability to use multiple CPU nodes for a single replica. Section~\ref{discussion} concludes with a summary of existing REMD frameworks and an analysis of RepEx, relative to other frameworks.

\section{Landscape of REMD simulations} \label{remd.landscape}

In this section we motivate the need for asynchronous REMD simulations and briefly discuss several novel software frameworks targeted at large-scale multi-dimensional RE simulations.  We conclude by capturing prevailing development trends and identify a gap in the current state-of-the-art.

\subsection{Asynchronous Replica Exchange} \label{async}

\begin{figure}[ht!]
  \centering
  \subfigure[]{
  \includegraphics[width=0.45\textwidth]{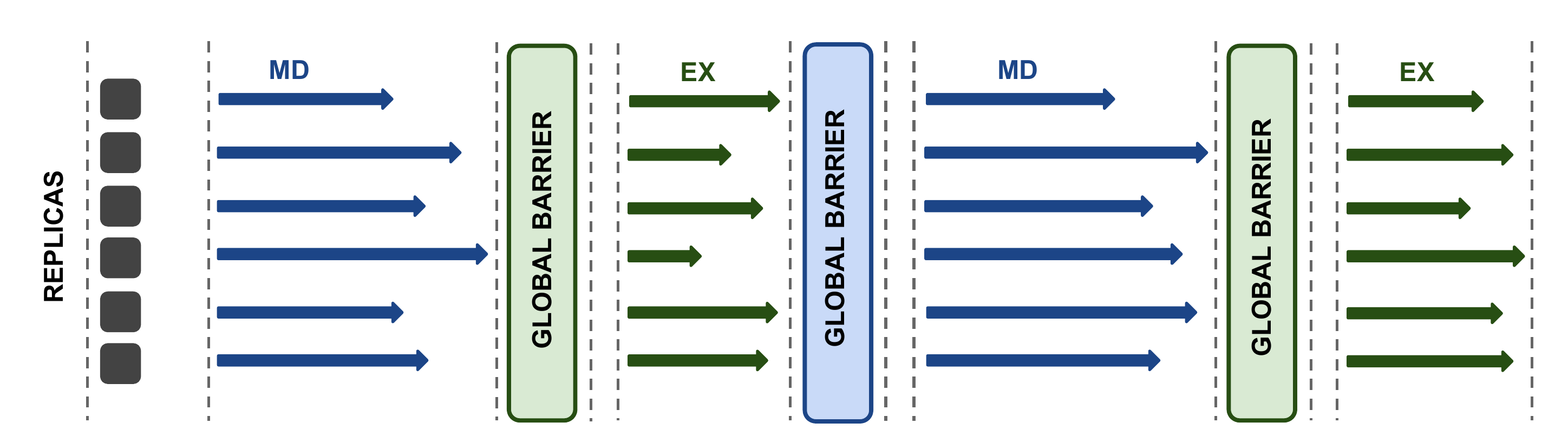}
  }
  \hfill
  \subfigure[]{
  \includegraphics[width=0.4\textwidth]{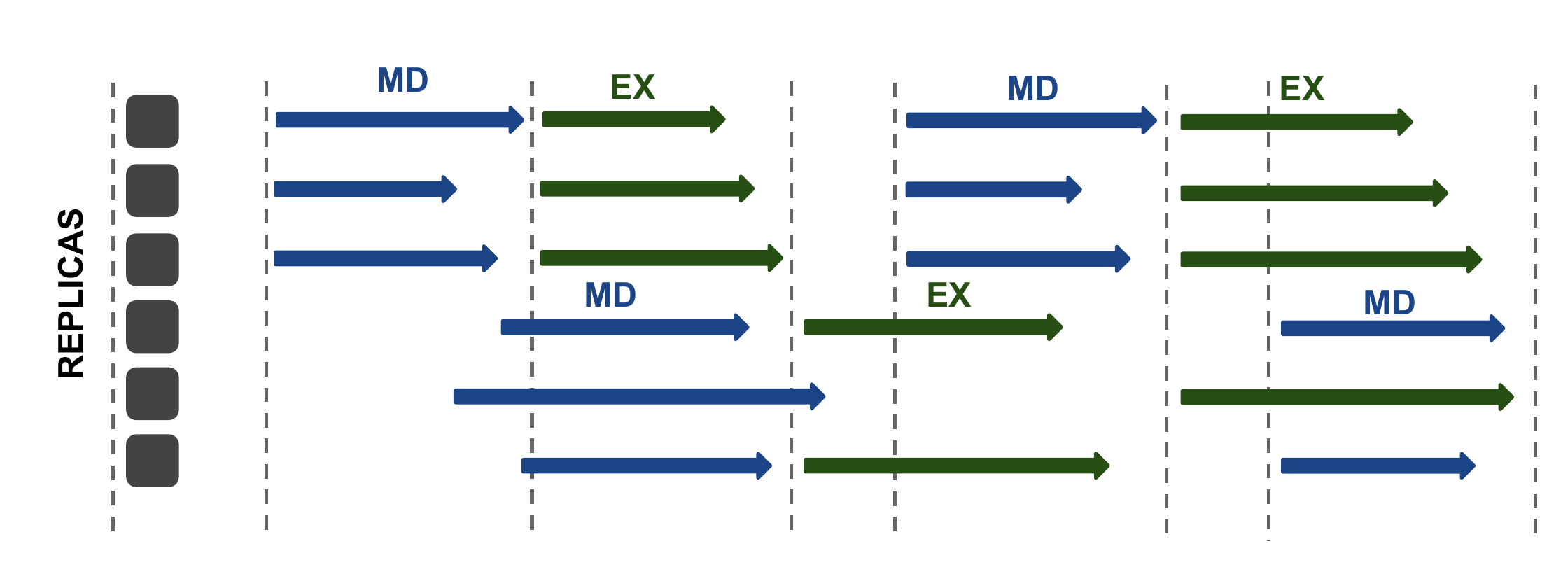}
  }
  \caption{\small{Schematic representation of Replica Exchange Patterns: (a) Synchronous (b) Asynchronous. x axis represents time. Gray squares represent replicas, blue arrows MD phase propagation and green arrows exchange phase propagation. In synchronous pattern, after both MD and exchange phase is present global barrier. In this figure, for synchronous pattern, both MD and exchange are propagated concurrently but this is not a requirement of this pattern. In asynchronous pattern there is no barrier - MD and exchange can be propagated concurrently, meaning while some replicas run MD other replicas might be running exchange.}  }
  \label{fig:re_patterns}
\end{figure}

RE algorithms have historically been synchronous, viz., there is a global barrier between the simulation and exchange phases (Figure~\ref{fig:re_patterns} (a)).  Asynchronous RE (Figure~\ref{fig:re_patterns} (b)) refers to the scenario when replicas can be in different phases. For example, a subset of replicas might be exchanging while some replicas might still be in simulation phase. In other words, the global synchronization of regular RE is relaxed.  Asynchronous RE has the following scientific advantages:

\textbf{Facilitates adaptive sampling.} There are cases, where some replicas have already produced sufficient info and are no longer needed. For example, replicas simulating configuration space with very low probability may not need high accuracy hence only relatively small amount of sampling is required.  Consequently these replicas should be terminated and their computational resource should be released.  On the other hand, in the midst of simulations, new replicas may need to be created to cover the regions where more sampling is necessary.  Obviously asynchronous algorithms are needed in such cases.

\textbf{Enables integration of heterogeneous simulations.} Nowadays multi-scale molecular
simulations may consist of very different levels of theories hence different replicas may
have significant differences in performance. For example, quantum mechanics calculations usually
are slower than classical molecular dynamics simulations. As a result, it is desired to have
asynchronous RE algorithms to handle simulations with large mismatch in performance.

\textbf{Handles fault-tolerance.} Large-scale RE simulations, are more susceptive to both 
hardware and software failures, which result in failures of individual replicas. Hence it is
necessary to recover from such failures and continue simulation. Due to the
nature of asynchronous algorithms, recovery time is significantly reduced compared to a
synchronous RE, where in case of a failure all other replicas must wait at the
barrier for a restarted replica.

\textbf{Manages load-balance with fluctuation of available resources.} Multi-dimensional RE
simulations may require very large numbers of replicas, which could be larger than the
available number of CPUs. In addition, both the number of running replicas and
availability of a resource could change during simulation. Traditional synchronous algorithms are
not capable to handle such cases. Asynchronous algorithms are needed to execute replicas at
different time so that simulations of all replicas can be performed.

\subsection{Related Work} \label{related.work}

In this section we focus on frameworks designed to implement RE algorithm outside of the MD engine; we defer a discussion of REMD simulations using molecular simulation software packages with integrated RE capability till later. We start with CHARMM based implementation for 2D REMD, then we introduce Multiple Copy Algorithm (MCA) implementation with NAMD engine and finally we discuss implementations of asynchronous Replica Exchange.

{\bf REPDSTR module of the CHARMM:} 
Ref. ~\cite{Jiang_JChemTheoryComput_2012_v8_p4672} presents an implementation of a 2D US/H-REMD method, implemented in REPDSTR module of the CHARMM~\cite{charmm}. REPDSTR uses an MPI level parallel/parallel mode where to each replica are assigned multiple MPI processes and dedicated I/O routines. To improve sampling efficiency, exchange attempts are performed alternatively along the two dimensions. 

Implementation was tested on IBM Blue Gene/P supercomputer using the binding of calcium ions to the small protein Calbindin $D_9k$. Obtained results show that 2D US/H-REMD significantly improves the configurational sampling for biological potential of mean force (PMF) calculations and as a result facilitates convergence of the simulation.

Strong scaling performance of 2D US/H-REMD, involving 4096 replicas and utilizing up to 131072 CPUs was presented, demonstrating nearly linear scaling.

{\bf MCA implementation with NAMD:} A Charm++ based implementation designed to run MCA was presented in ~\cite{jiang2014}. It is tightly bound to the NAMD simulation engine. Charm++ is used to concurrently run multiple NAMD instances, which are exchanging messages via point-to-point communication functions of Tcl scripting interface. Tcl scripting enables users to implement REMD algorithms without modifying the source code.  Authors demonstrated strong scaling behavior of the swarms-of-trajectories string method implementation using the full-length c-Src kinase system utilizing up to 524288 cores on Blue Gene/Q supercomputer. In addition, results of T-REMD simulations of peptide acetyl-$(AAQAA)_3$-amide ~\cite{shalongo} in TIP3 solvent on Blue Gene/Q utilizing up to 32768 cores were presented.

{\bf Asynchronous approaches:} Ref.~\cite{2013-xsede-cdi,ct500776j} presented ASyncRE package, developed to perform large-scale asynchronous REMD simulations on HPC systems. ASyncRE has an emphasis on asynchronous RE. Package supports Amber~\cite{Amber} and IMPACT~\cite{impact2005} MD engines. It implements two REMD algorithms, namely multi-dimensional RE umbrella sampling with Amber and BEDAM $\lambda$ RE alchemical binding free energy calculations with the IMPACT. AsyncRE uses a similar runtime system as RepEx, is capable of launching more replicas than there are CPU cores allocated and is fault tolerant: failure of a single (or multiple) replicas does not result in failure of a whole simulation. If needed, new replicas can be launched to compensate for a failed ones.

Ref.~\cite{Xia_JComputChem_2015_v36_p1772} introduced another REMD package targeted at asynchronous RE, optimized for volunteer computing resources. Package can be used on HPC clusters as well. It is customized for IMPACT as MD simulation engine and supports both 1D and 2D REMD simulations.
Distinctive features are: fault tolerance, the ability to use a dynamic pool of resources and to use less CPU cores than replicas. Exchange phase is performed on coordination server, meaning that output data must be moved from target resource to coordination server.

{\bf Summary of Related Work:} As can be seen, there are multiple existing software packages designed to perform large-scale REMD simulations. The packages discussed, represent a small fraction of available tools, they still captured the primary important points: A significant number of tools, support only a single MD simulations engine and are designed in a way which makes it very difficult to substitute MD simulation engines. There are many scientific reasons the community has supported distinct multiple MD simulation engines, but the tight binding of replica-exchange methods to a particular engine raises a barrier for the uptake of new simulation codes. The advantage of close integration is admittedly scalability, although this comes at the price of the lack of generality of both the replica-exchange methods and simulation engine supported. For example, even though the Charm++ based framework has impressive scalability, a typical user can use it for only the replica-exchange methods that it supports. Importance of execution options and support for both asynchronous and synchronous RE is often underestimated.

Historically, tight integration has been prevailed due to the perception that performance trumps all other features. There are emerging examples of important biomolecular problems however, that involve multi-state equilibria, and for which the interpretation of experiments requires scanning control variables such as temperature, ionic conditions, and pH in addition to geometrical or Hamiltonian order parameters\cite{Bergonzo_JChemTheoryComput_2014_v10_p492}.
These applications have the added challenge that sampling along the space of the order parameters needs to be statistically converged at all points. Here, the REMD method offers the added advantage that equilibrium between simulations is enforced through the exchange sampling. An illustrative example is the ''problem space'' associated with biocatalysis whereby conformational equilibria, metal ion binding and protonation events lead to an active state that is able to catalyze the chemical steps of the reaction~\cite{Panteva_BookChap_MultiscaleRNAEnzym_2015_v553_p335}.  Thus, these applications require not only the elucidation of the free energy landscape of the chemical reaction itself~\cite{Ensing_AccChemRes_2006_v39_p73, Vanden-Eijnden_JComputChem_2009_v30_p1737},
but also the characterization of the probability of finding the system in the catalytically active state as a function of system variables~\cite{Dissanayake_Biochemistry_2015_v54_p1307}. To address these novel applications and scenarios, a flexible and efficient multi-dimensional REMD framework is required, that can be used for both system control variables and generalized coordinates. Currently there is no REMD framework capable of providing the required flexibility in composing the range of RE methods with MD engines as needed while providing adequate performance.

\section{RepEx: A Framework for Replica Exchange} \label{repex.framework}

We outline the requirements of a framework that would support the needs of scientific problems~\cite{Ensing_AccChemRes_2006_v39_p73, Vanden-Eijnden_JComputChem_2009_v30_p1737, Dissanayake_Biochemistry_2015_v54_p1307} that are not being met currently. We then introduce {\bf RepEx} -- a framework designed to meet these requirements and discuss its implementation.

\subsection{REMD Requirements} \label{requirements}   

In this sub-section, we motivate and define requirements of a REMD simulations software. We describe three types of requirements: functional, performance (scalability) and usability.  We identified the following functional requirements:

\textbf{Generality} is a requirement to maximize a range of replica-exchange methodologies as well as MD engines. A general purpose framework should support: (i) different types of exchange parameters, (ii) multiple exchange parameters in a single REMD simulation, and (iii) multiple MD engines.  A corollary of this requirement, is the decoupling of advances in replica-exchange methodology to MD engines, and thus the potential for broader (greater number of REMD applications) and deeper impact (enable new research opportunities).

\textbf{Execution flexibility} arises from the need to decouple the number of CPU cores from the number of replicas. Alternatively, the ability to set-up a REMD simulation with a desired number of replicas, independent of the number of CPU cores available at a given instance of time.  For reasons ranging from a queue waiting time to limitations in number of allocatable CPU cores on a given cluster, it should be possible to use as many or as few CPU cores as needed, irrespective of the number of replicas. The same principle also applies to individual replicas: support for both single-core and multi-core replicas should be provided. Currently {\bf all} REMD frameworks require at least as many CPU cores as replicas; further more, the number of replicas that are actively simulated is fixed and statically determined. 

\textbf{Synchronization.} In order to enable a wider range of REMD simulations support for asynchronous RE is required without loss of generality or execution flexibility. 


\textbf{Interoperability.} Most REMD simulations are executed on supercomputers which vary in scheduling systems, middleware and software environment. In order to support community production grade science, an REMD framework should work on multiple high-end machines as well as small HPC clusters, while retaining functionality and performance.

The above functional requirements have to be balanced with the following performance requirements:

\textbf{Scalability with the number of replicas.} To obtain high sampling quality, REMD simulations should support the ability to run a large number of replicas. Furthermore, given that the number of replicas needed in an REMD simulation scales as $\approx N^d$, where $d$ is the dimensionality (of exchange), the need to support a large number of replicas is greater when applied to multi-dimensional simulations. Achieving good scalability for REMD frameworks is a challenging task, especially when preserving the four functional requirements outlined above.

\textbf{Scalability with the number of CPU cores.}  Whereas the primary performance metric is the scalable execution of a large number of replicas, it is often the case that each replica is multi-node (in Ref.~\cite{swadling2015structure}, each replica was 768 cores); multi-node replicas are important in order to simulate large physical systems.  Any framework should provide scalability in the number of replicas simulated and the number of CPU cores utilized.

Last but not least, we briefly discuss usability considerations specific to REMD simulations.
  
\textbf{Usability.} Relative to the simulation phase, the exchange phase is significantly more complex. Thus, not only should a framework for REMD separate the logic of the exchange mechanisms from the simulation mechanisms, it should not expose the complexity of exchange mechanism, should be automated as much as possible and must be fully specified by configuration files. Definition of configuration files should be intuitive and should include a minimal set of parameters.

\subsection{Design} \label{design} 

To satisfy the requirements outlined in the previous sub-section, we discuss the three concepts underpinning the unique design of RepEx: 
\begin{compactitem}
  \item \textbf{Replica Exchange Pattern:} explicit support for synchronization patterns between simulation and exchange phases.
  \item \textbf{Pilot-Job system:} a multi-stage mechanism for workload execution via the use of an initial placeholder job (the ``pilot'') and thus dynamically allocating computational resources for replicas.
  \item \textbf{Flexible Execution Mode:} The ability to execute different patterns and number of replicas independent of the underlying resources available, i.e., flexible spatial and temporal mapping of workload (tasks) to the allocated CPUs.
\end{compactitem}

\subsubsection{Replica Exchange patterns} \label{re.patterns}
 
The RepEx framework captures the distinction between different synchronization scenarios using two RE patterns and exposes them to end-users, independent of MD simulation engine and the resources available.

{\bf Synchronous RE Pattern:} Synchronous RE pattern depicted in Figure~\ref{fig:re_patterns} (a), corresponds to the scenario where all replicas must finish simulation phase, before any of the replicas can transition to the exchange phase. There is a global synchronization barrier, which forces the replicas arriving at the barrier early to wait for the lagging replicas. Once all replicas are done in the simulation phase all of them transition to the exchange phase. This cycle is then repeated. The synchronous pattern is the conventional way of running REMD simulations, partly because of the implementation simplicity.

{\bf Asynchronous RE Pattern:} Asynchronous RE Pattern, shown in Figure~\ref{fig:re_patterns} (b), does not have a global synchronization barrier between simulation and exchange phase. While some replicas are in the simulation phase, others might be in the exchange phase. Based on some criterion, a subset of replicas transition into the exchange phase, while other replicas continue in the simulation phase. Selection of replicas that will transition may be based on a FIFO principle, e.g. first N replicas transition into an exchange phase. Alternatively, only replicas which have finished a predefined number of simulation time-steps (2 ps) during some real time interval (1 minute) transition into exchange phase.

\subsubsection{Pilot-Job systems} \label{pilot.job.systems}

The Pilot-Job concept was originally introduced to reduce queue waiting times for workloads on distributed clusters.  Pilot-Job systems have been generalized~\cite{review_pilotreview_2013} to provide a variety of capabilities, but the two most important are: management of dynamically varying resources and execution of dynamic workloads. A Pilot-Job system supports the execution of workloads with multiple, heterogeneous and dependent tasks\cite{review_radicalpilot_2015}. When the placeholder job (pilot) becomes active, it can start executing tasks on acquired computational resources. Resources are available for a duration specified in a job description. Tasks can be submitted for execution before or after the pilot becomes active. Integration of a Pilot-job system in our design enables different RE patterns and number of replicas, independent of the resources available, i.e., supports flexible execution modes, which we now discuss.

\subsubsection{Flexible Execution Modes} \label{fem}

Decoupling the workload size from the available resources requires: (i) the details of workload be kept separate from the details of resources --- type, quantity and availability, and (ii) the ability to execute a workload of a given size (say N replicas) independent of the specific resources available.  As alluded to, Pilot-job systems enable the former; we now discuss how the the pilot abstraction enables the latter.

Depending upon the relative size of the resources available (R) to the size of simulations (S = number of replicas x resource requirement of each replica), REMD simulations are executed differently. Thus there are two Execution Modes: when R > S (Execution Mode I), and when R < S (Execution Mode II), each of which can be used with any of the two RE patterns.

\begin{figure}
  \centering
  \includegraphics[width=0.45\textwidth]{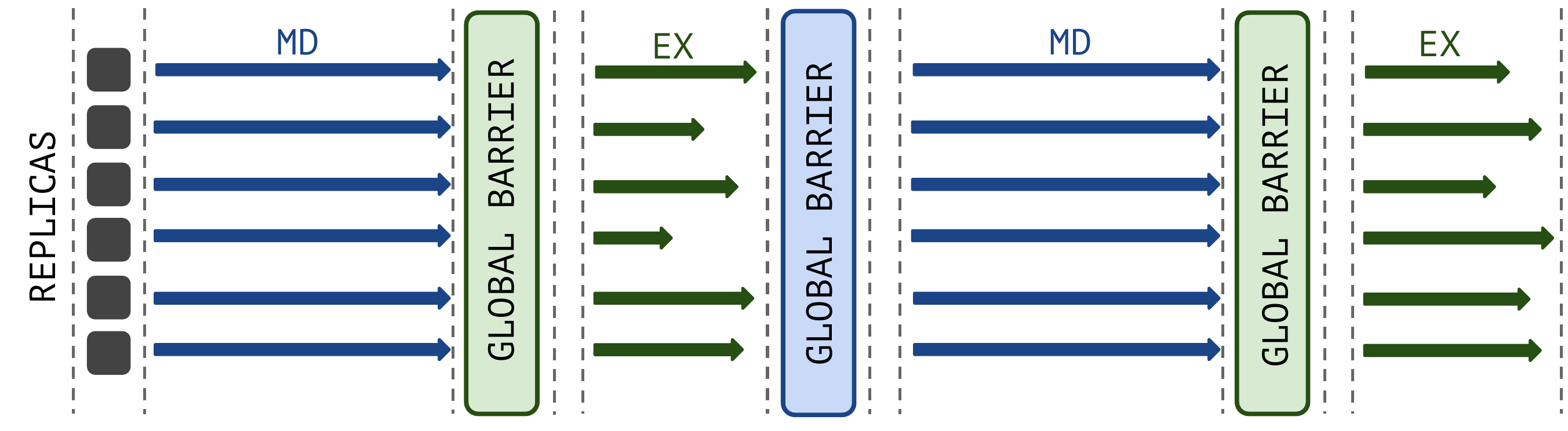}
  \caption{\small{Schematic representation of Execution Mode I. On the x axis is time. Gray squares represent replicas, blue arrows MD phase propagation and green arrows exchange phase propagation. Both MD and exchange phase for all replicas are performed concurrently. After MD and exchange phase is placed global barrier, ensuring that all replicas enter next phase simultaneously.}}
  \label{fig:exec.mode.1}
\end{figure}

{\bf Execution Mode I:} In Execution Mode I the number of allocated CPU cores satisfies execution requirements of all replicas at a given instant of time. For example, if each replica requires a single CPU core to run, in this mode enough cores are allocated to run all replicas concurrently. Figure~\ref{fig:exec.mode.1} illustrates capturing Execution Mode I in the context of a Synchronous RE pattern.

\begin{figure}
  \centering
  \includegraphics[width=0.45\textwidth]{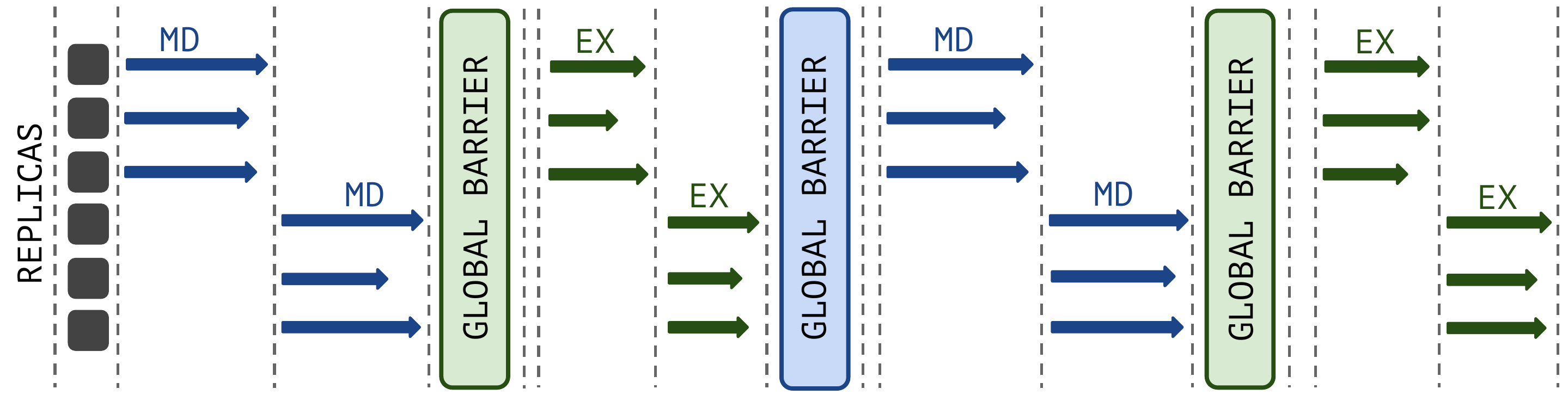}
  \caption{\small{Schematic representation of Execution Mode II. On the x axis is time. Gray squares represent replicas, blue arrows MD phase propagation and green arrows exchange phase propagation. Replicas don't propagate MD and exchange phase concurrently. Batch size for each phase is determined by the number of CPU cores allocated.  A global synchronization barrier is present after both MD and exchange phase, ensuring that all replicas enter next phase simultaneously.}}
  \label{fig:exec-mode-2}
\end{figure}

{\bf Execution Mode II:} Execution Mode II supports the scenario when there are not enough CPU cores to run all replicas concurrently. The ratio of cores to replicas is a user defined variable, but typically is a term of a geometric series, e.g. $\frac{1}{2}$, $\frac{1}{4}$, $\frac{1}{8}$, $\frac{1}{16}$. As a result, only a fraction of replicas can propagate simulation or exchange phase concurrently. A schematic representation of Execution Mode II is illustrated in Figure~\ref{fig:exec-mode-2}; for simplicity we depict the synchronous RE Pattern, but Execution Mode II can be used with any of the two available RE Patterns.

While users are given an option to select an execution mode, exact execution details are determined by execution management module of RepEx.  The Execution Mode abstraction hides the gory details of the execution, which differ based upon the relative values of R and S. The implementation of Execution Modes differ in the spatial and temporal mapping of workload (tasks) to the allocated CPUs. Specifically, they differ in: 
\begin{compactitem}
  \item Order and level of concurrency for task execution 
  \item Number of Pilots used for a given simulation
  \item Number of concurrently used target HPC resources
\end{compactitem}
 
Execution Mode is a subset of execution options decoupling simulation requirements from the resource availability and enabling flexible usage of allocated HPC resources.  A user should be able to switch between available Execution Modes without any refactoring.  In addition to providing conceptual simplicity by hiding details of the execution, Execution Modes provides an important practical functionality: it permits the study of systems not otherwise possible thanks to the ability to launch more replicas then there are allocatable CPU cores on a target HPC cluster. This might be particularly useful when a user has access to small HPC clusters, but is interested in running REMD simulations involving large number of replicas. For example, a user can assign as many cores to each replica as needed, or if only a small HPC cluster comprising 128 cores is available, user still can perform a simulation involving 10000 replicas.

\subsection{Implementation} \label{implementation}

RepEx is an open-source package released under the MIT license. RepEx source code and documentation are available at ~\cite{repex-code}. For execution of its workloads RepEx relies on a concept of task-level parallelism, which is enabled by the RADICAL-Pilot system.

Currently RepEx supports two MD simulation engines (Amber and NAMD) and three exchange parameters (temperature (T-REMD), biasing potential (U-REMD) and salt concentration (S-REMD)).  The individual exchange parameters can be combined into multi-dimensional REMD with arbitrary ordering and number of dimensions. RepEx can be extended to support other MD simulation engines, exchange parameters, REMD types and execution modes.  At the core of RepEx are three modules:

{\bf Execution Management Modules (EMM):} EMM enables a separation of execution details (viz., resource management and workload configuration) from the simulation using different MD engines. Most of the resource management (RP API) calls are performed in EMM, such as instantiation of a pilot and its launching on a target resource, via a translation of the user requirements. A single EMM is used for all 1D-REMD (or 3D-REMD) simulation types. Encapsulation of synchronization routines by EMM, allows to fully specify synchronous or asynchronous RE by a single EMM. Finally, EMM is MD engine independent.

{\bf Application Management Modules (AMM):} AMM support {\it generality} by managing exchange parameters, input parameters, simulation input/output files and file movement patterns. AMM is the first module which is instantiated during execution. AMM is then used to instantiate replica objects according to the {\it simulation input file}. AMM is specific to a particular MD engine, since input/output files and arguments for each MD engine are different. Thus AMM performs the translation of user requirements specified in simulation input file.  After pilot becomes active, in EMM is entered main simulation loop, where is called a method of AMM to prepare tasks (RP's \textbf{Compute Units}) for both MD and exchange phase. These tasks are then passed to EMM for execution.
    
{\bf Remote Application Modules (RAM):} RAM is responsible for creation of individual input files for replicas, reading data from simulation output files and performing exchange procedures. Unlike EMM and AMM which are client side, these modules execute on HPC cluster.  RAMs provide {\it scalability} and {\it execution flexibility}.

\subsection{Validation} \label{validation}

\begin{figure}
   \centering
   \includegraphics[width=0.5\textwidth]{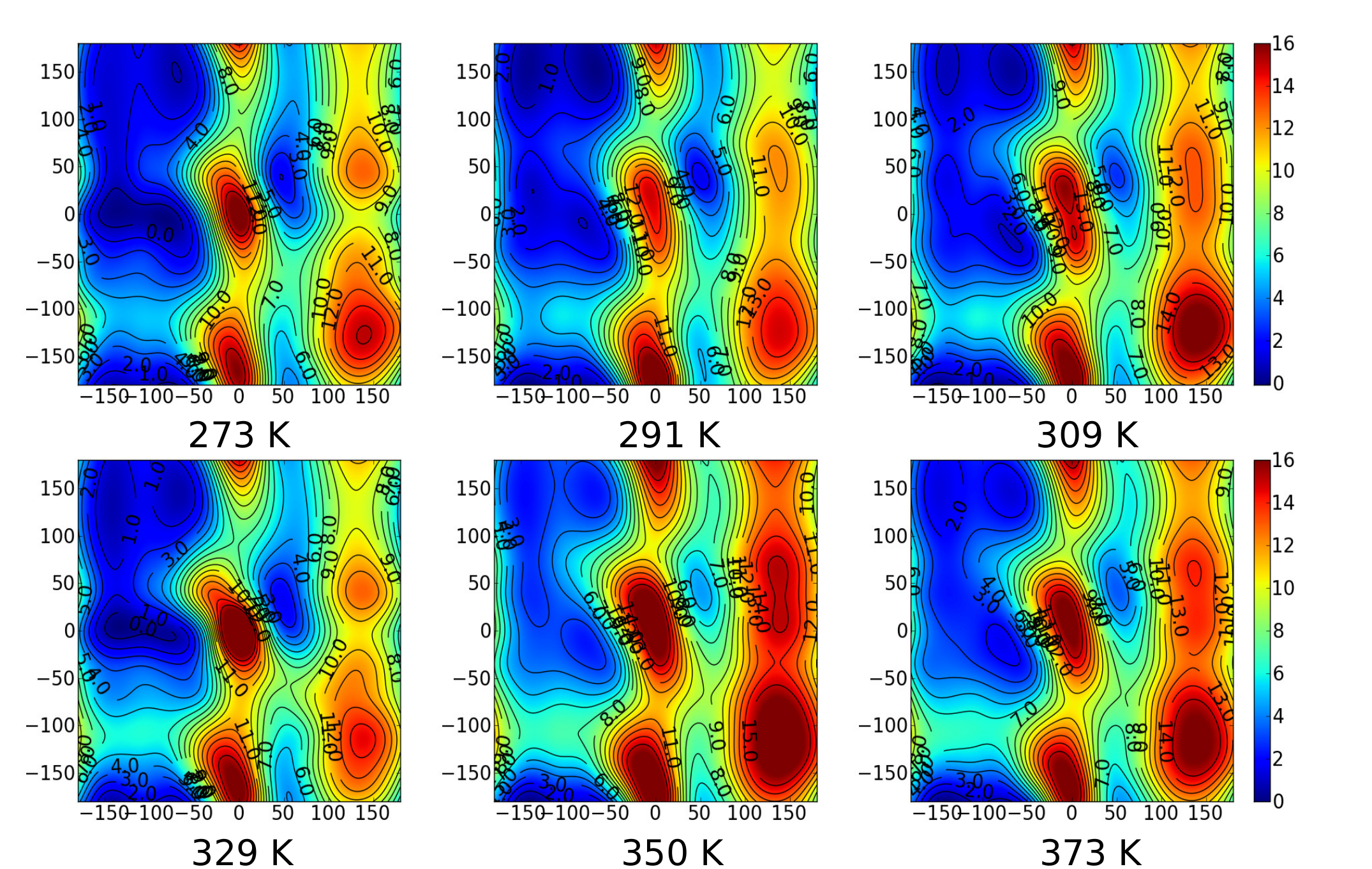}
   \caption{\small{Free energy profile of alanine dipeptide backbone torsion at 6 different temperatures. In all 6 subplots, the x and y axes correspond to $\phi$ and $\psi$ torsion angles, respectively. The range of energies is from 0 kcal/mol to 16 kcal/mol while each level in the contour corresponds to a 1 kcal/mol increment.}}{\label{fig:energy_maps}}
\end{figure}

3D-REMD was performed using order parameters of temperature, and umbrella sampling in the $\phi$ and $\psi$ torsion angles (as shown in Figure~\ref{fig:energy_maps}). In the T dimension, 6 windows were chosen from 273K to 373K by geometrical progression. In both U dimensions, 8 windows were chosen uniformly between 0$^{\circ}$ and 360$^{\circ}$ where each window corresponds to a harmonic restraint centered on it with a force constant of 0.02 kcal${\cdot}$mol$^{-1}{\cdot}{\textrm{degree}}^{-2}$. The total number of replicas is therefore 6$\times$8$\times$8=384. Each replica was previously equilibrated for >1 ns. In the production run, we set the exchange attempt interval (cycle) to be 20000 steps (20 ps) and in a 15-hour run with 400 cores (25 nodes) on the Stampede~\cite{xsede} cluster, the simulation finished 90 cycles (1.8 ns). The acceptance ratios of exchange attempts are approximately 3 \% for T dimension and 25 \% for U dimensions. Free energy profiles were then generated from the last 1 ns of production data using the maximum likelihood approach implemented in the vFEP package~\cite{Lee_JChemTheoryComput_2013_v9_p153,Lee_JChemTheoryComput_2014_v10_p24}.

\section{Experiments} \label{experiments}

Having validated both the design and implementation of RepEx, in this section we discuss a series of experiments used to demonstrate the unique functional capabilities and characterize its performance.

For all experiments we measure and plot average REMD simulation cycle time, which is average of 4 simulation cycles. Experiments were performed using alanine dipeptide (Ace-Ala-Nme) solvated by water molecules on Extreme Science and Engineering Discovery Environment ~\cite{xsede} (XSEDE) allocated systems: Stampede and SuperMIC~\cite{xsede}. For all experiments, unless stated otherwise the Synchronous RE pattern was used.

Simulation cycle time is defined as:   
\begin{equation}
T_{c} = T_{MD} + T_{EX} + T_{data} + T_{RepEx-over} + T_{RP-over}
\end{equation}
where:
\begin{compactitem}
  \item $T_{MD}$ - MD simulation time, time to perform X simulation time-steps  
  \item $T_{EX}$ - Exchange time. Time for calculations required to determine exchange partners
  \item $T_{data}$ - Data time. Time to perform data movement procedures, which are mostly remote-to-remote. For example, Amber's .mdinfo files to "staging area" which is accessible by subsequent tasks
  \item $T_{RepEx-over}$ - RepEx overhead. Time to prepare tasks for execution and time to perform local RepEx method calls
  \item $T_{RP-over}$ - RP overhead. Time required for task launching on a target resource and time for internal RP communication
\end{compactitem}

For M-REMD simulations, $T_c$ is comprised of the 1-D cycle time for each dimension, since simulations are performed only in one dimension at any given instant of time.

We calculate weak scaling efficiency as:
\begin{equation}
E_{w} = \frac{T_{1}}{T_N} \times 100 \% 
\end{equation}
where:
\begin{compactitem}
  \item $T_{1}$ - time to complete simulation cycle involving minimal number of replicas $R_{min}$ with number of CPU cores equal to the number of replicas, e.g. 8 replicas on 8 CPUs
  \item $T_{N}$ - time to complete simulation cycle involving $N$ replicas with $N$ CPU cores
\end{compactitem}
We calculate strong scaling efficiency as:
\begin{equation}
E_{w} = \frac{T_{1}}{N \times T_N} \times 100 \% 
\end{equation}
where:
\begin{compactitem}
  \item $T_{1}$ - time to complete simulation cycle involving $N$ replicas with minimal number of CPU cores $N_{min}$, e.g. 1024 replicas on 8 CPUs
  \item $T_{N}$ - time to complete simulation cycle involving $N$ replicas with $M$ CPU cores, where $N_{min} < M$
\end{compactitem}

Results obtained in Section~\ref{experiments} can be reproduced by following instructions at ~\cite{repex-experiments}. All experiments were performed with RADICAL Pilot version 0.35. The latest version of RP is 0.38, and thanks to various optimizations it is capable of substantially better performance than the version we have used for our experiments. These optimizations however, only alter the RP overhead timings (as well as data timings) presented in this section and will have minimal impact on the overall performance characterization of RepEx.

\subsection{Characterization of Overheads} \label{over}

There are three factors which contribute to the $T_{c}$ as a result of design decisions we have made. These factors are: data time, RepEx overhead and RP overhead. In this subsection we summarize how these factors influence the $T_{c}$.

\begin{figure}[ht!]
  \centering
  \includegraphics[width=0.4\textwidth]{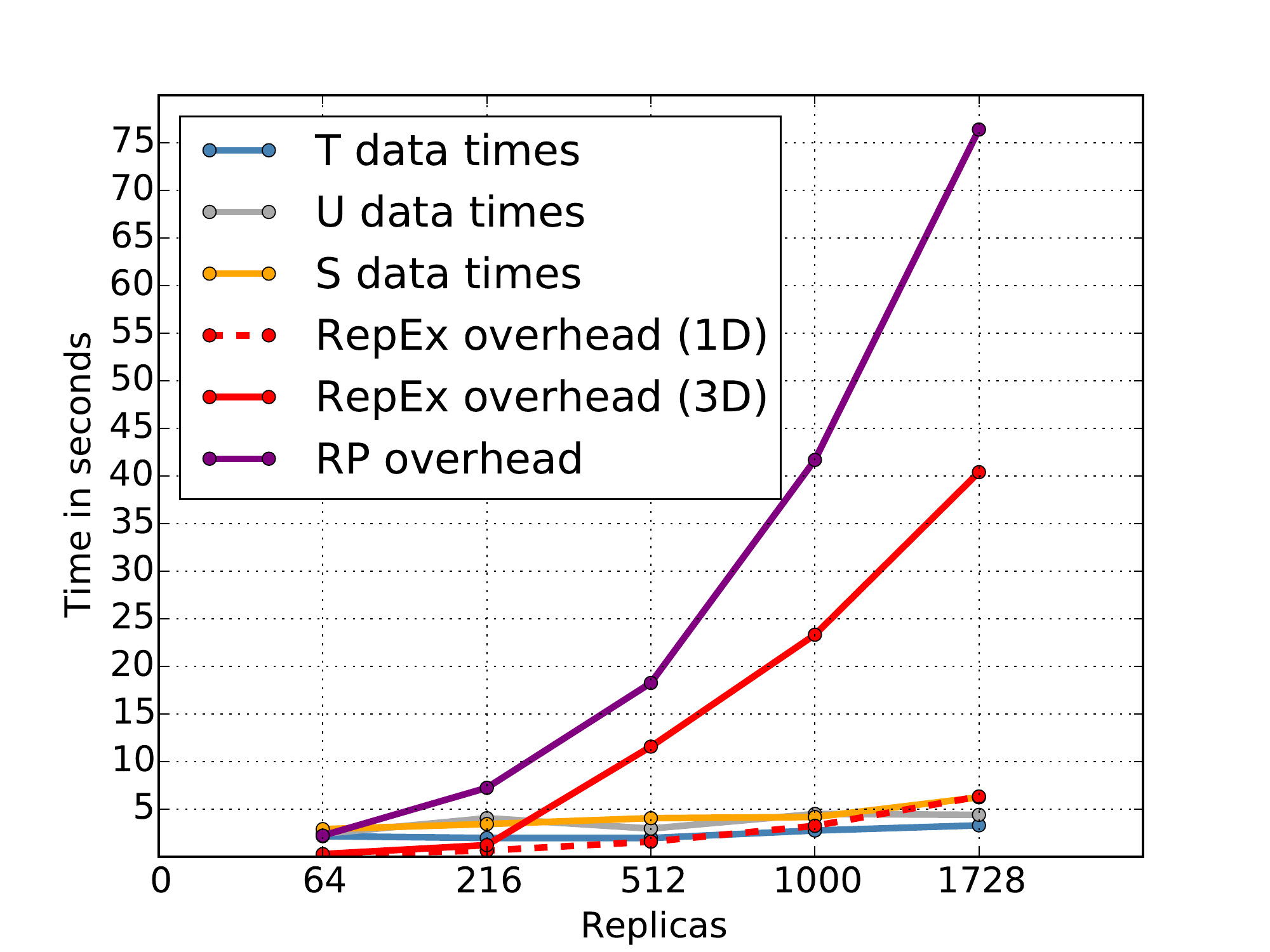}
  \caption{\small{Characterization of overheads: Data times, RepEx overhead and RP overhead. }
  }
  \label{fig:overheads}
\end{figure}

Figure~\ref{fig:overheads} presents the values of data times, RepEx overheads and RP overheads for simulation runs involving 64, 216, 512, 1000 and 1728 replicas on SuperMIC. For all runs we use a single CPU core per replica and use Execution Mode I with synchronous RE pattern.

Values of data times depend on the exchange type, since data movement patterns differ for each exchange type. As depicted in Figure~\ref{fig:overheads}, data times for temperature exchange are shorter than for umbrella exchange and salt concentration. For all replica counts, data times are relatively small: longest data transfer time is 6.3 seconds. This is due to the fact, that majority of the transfers are happening within the cluster/resource. Consequently, data times change as a function of a target system, since largest contributing factor is performance of a parallel file system.

RepEx overhead depends on the total number of replicas and on simulation type. For all 1D simulations, values of RepEx overhead are nearly identical, since number of operations to perform task preparation is very similar.  RepEx overhead times for 3D simulations are longer, since there are more data associated with each replica, complexity of data structures is increased and more computations are performed during task preparation. 

RP overhead depends only on the number of replicas (tasks) launched concurrently. As we can see in Figure~\ref{fig:overheads}, RP overhead is proportional to the number of replicas.   

\subsection{Performance Characterization of 1D-REMD} \label{exp.1d-remd}

\begin{figure}[ht!]
  \centering
  \includegraphics[width=0.45\textwidth]{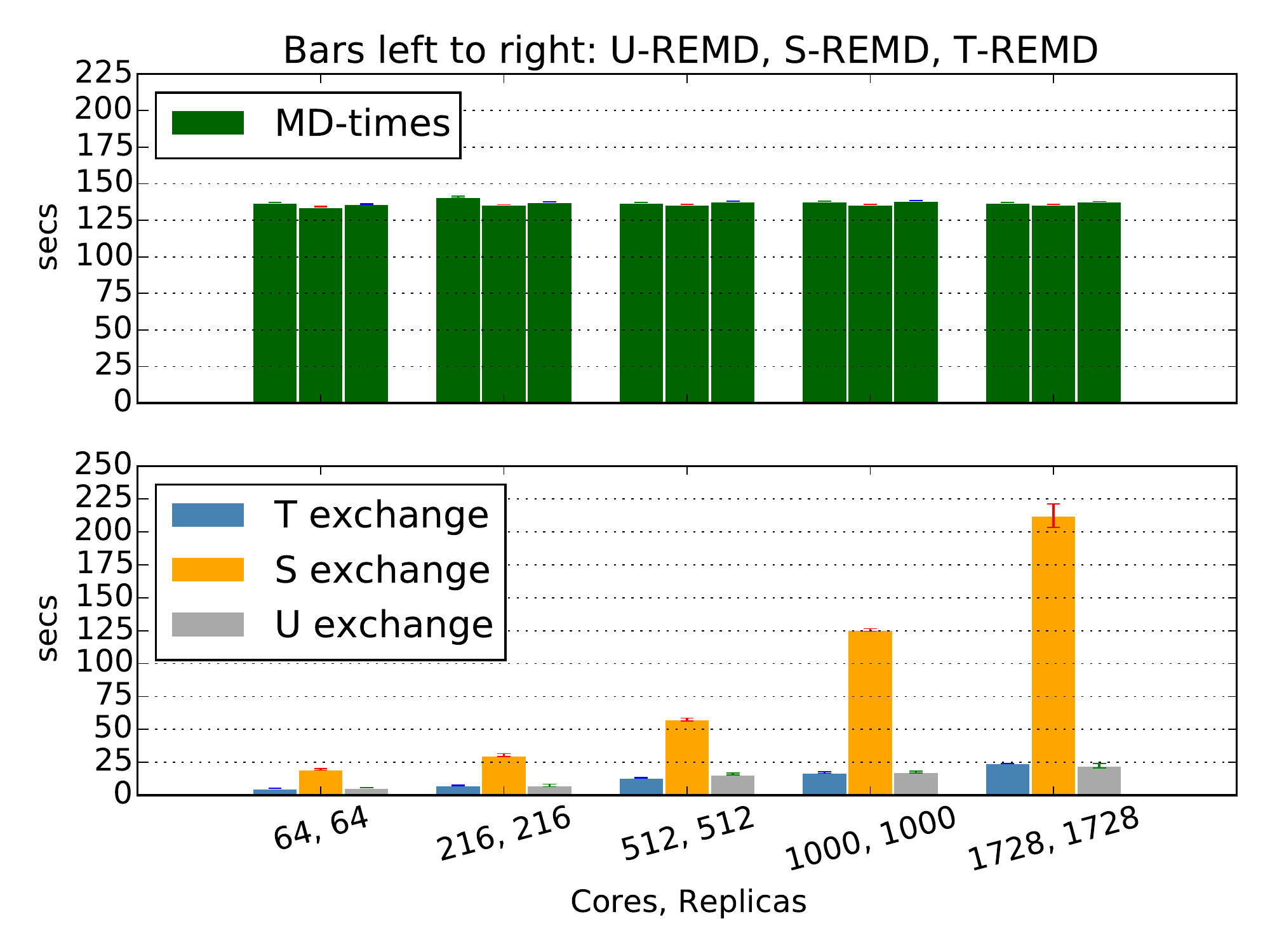}
  \caption{\small{One-dimensional REMD experiments with RepEx: weak scaling. Decomposition of average simulation cycle times $T_{c}$ (in seconds) into MD simulation time and exchange time for umbrella sampling, salt concentration and temperature exchange. For all simulation runs number of replicas is equal to the number of CPU cores and both vary from 64 to 1728. All simulation runs are performed on SuperMIC supercomputer. For all  runs are used single-core replicas.}  
  }
  \label{fig:1d_results}
\end{figure}

In this subsection we characterize performance of 1D REMD simulations with RepEx. For each of the three available 1D simulations: T-REMD, U-REMD and S-REMD we measure average cycle times. We perform simulation runs involving 64, 216, 512, 1000 and 1728 replicas in Execution Mode I. All runs are performed with a single CPU core per replica and \textbf{sander} as the Amber executable. We use alanine dipeptide solvated by water molecules comprising a total of 2881 atoms and perform 6000 simulation time-steps between exchanges. We perform all runs on SuperMIC supercomputer ~\cite{xsede}. Results of these experiments are presented in Figure~\ref{fig:1d_results}.

As we can see, for all three exchange types, the time to perform 6000 time-steps is nearly identical, as evidenced by the almost similar average heights of dark green bars in Figure~\ref{fig:1d_results} (139.6 seconds).

Next we discuss exchange timings for different exchange parameters, as seen in the lower panel of Figure~\ref{fig:1d_results}. Timings for temperature and umbrella exchange are similar and have a nearly linear growth rate. For both exchange types we use a single MPI task to perform an exchange. In case of U-REMD we have implemented a single point energy calculation internally. Despite the fact that U-REMD exchange is more involved, we don't see a significant difference in exchange timings between U-REMD and T-REMD. 

Due to the mathematical complexity, the single point energy calculation for S-REMD is calculated using Amber for each replica in each state.  This implies that for each replica, an additional task is required. Since we are using Amber's group files, this task requires at least as many CPU cores as there are potential exchange partners for each replica. Consequently, the exchange times for S-REMD are substantially longer, but nonetheless have a nearly linear growth rate.

\begin{figure}[ht!]
  \centering
  \includegraphics[width=0.3\textwidth]{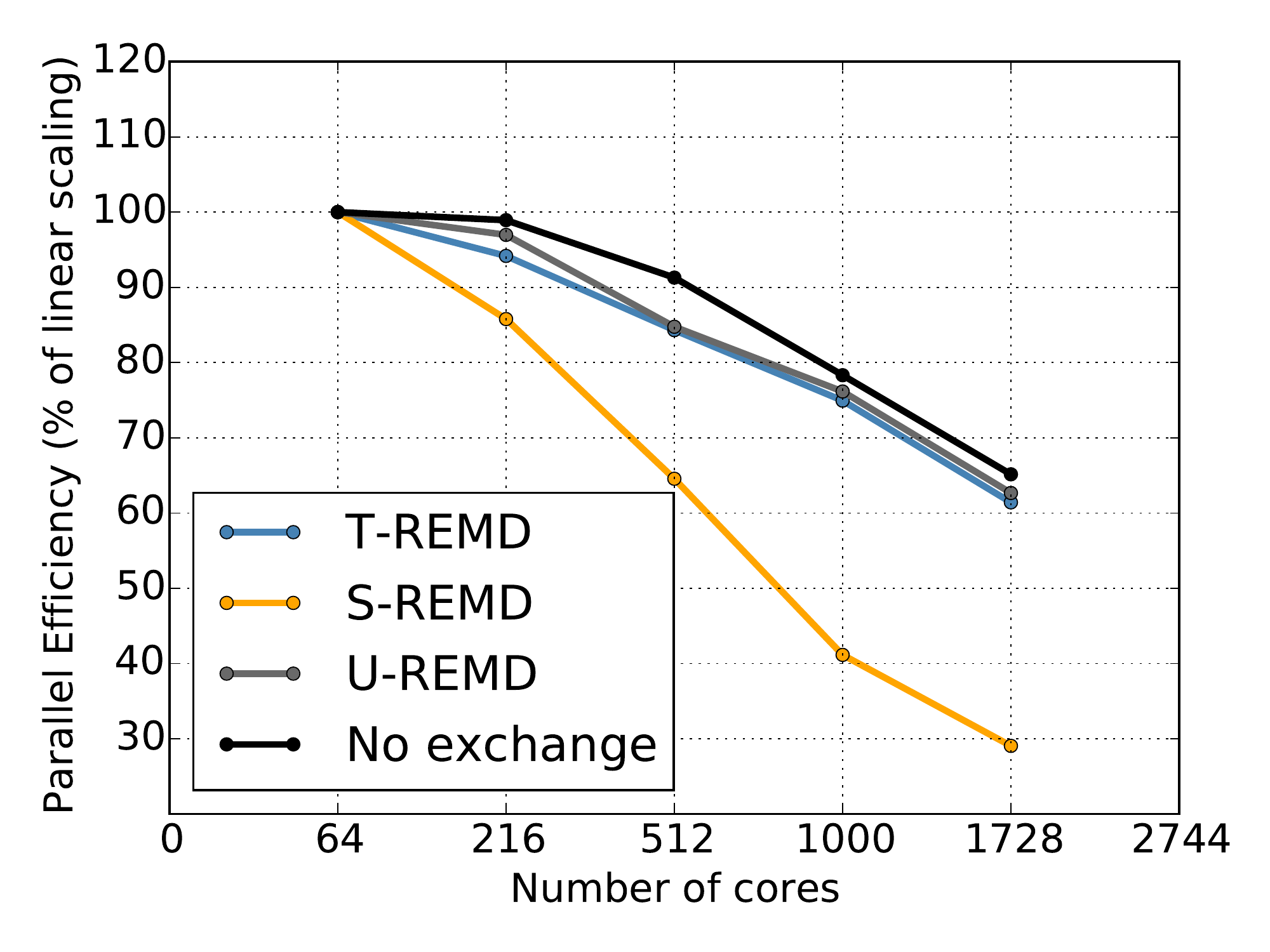}
  \caption{\small{Parallel Efficiency (\% of linear scaling) for Temperature Exchange REMD (1D), Salt Concentration REMD (1D) and Umbrella Sampling REMD (1D) using Amber MD engine on SuperMIC supercomputer.}
  }
  \label{fig:p_efficiency_1d}
\end{figure}

The parallel efficiency results for the 1D-REMD simulations are presented 
in Figure~\ref{fig:p_efficiency_1d}. We calculate parallel efficiency for the weak 
scaling scenario and use average cycle time for simulation with 64 cores as 
starting point, e.g. 100\% efficiency. We also present efficiency results for 
simulations without an exchange phase (black line). This quantifies the influence 
of exchanges on efficiency of 1D simulations. For all three exchange types we observe 
decrease in efficiency while increasing the number of cores. Efficiency values 
for T-REMD and U-REMD are similar and demonstrate linear behavior. Efficiency for S-REMD 
is lower. This is caused by specifics of exchange phase, discussed earlier.

\subsection{T-REMD with NAMD engine} \label{exp.namd.t-remd}

\begin{figure}[ht!]
\centering
\includegraphics[width=0.45\textwidth]{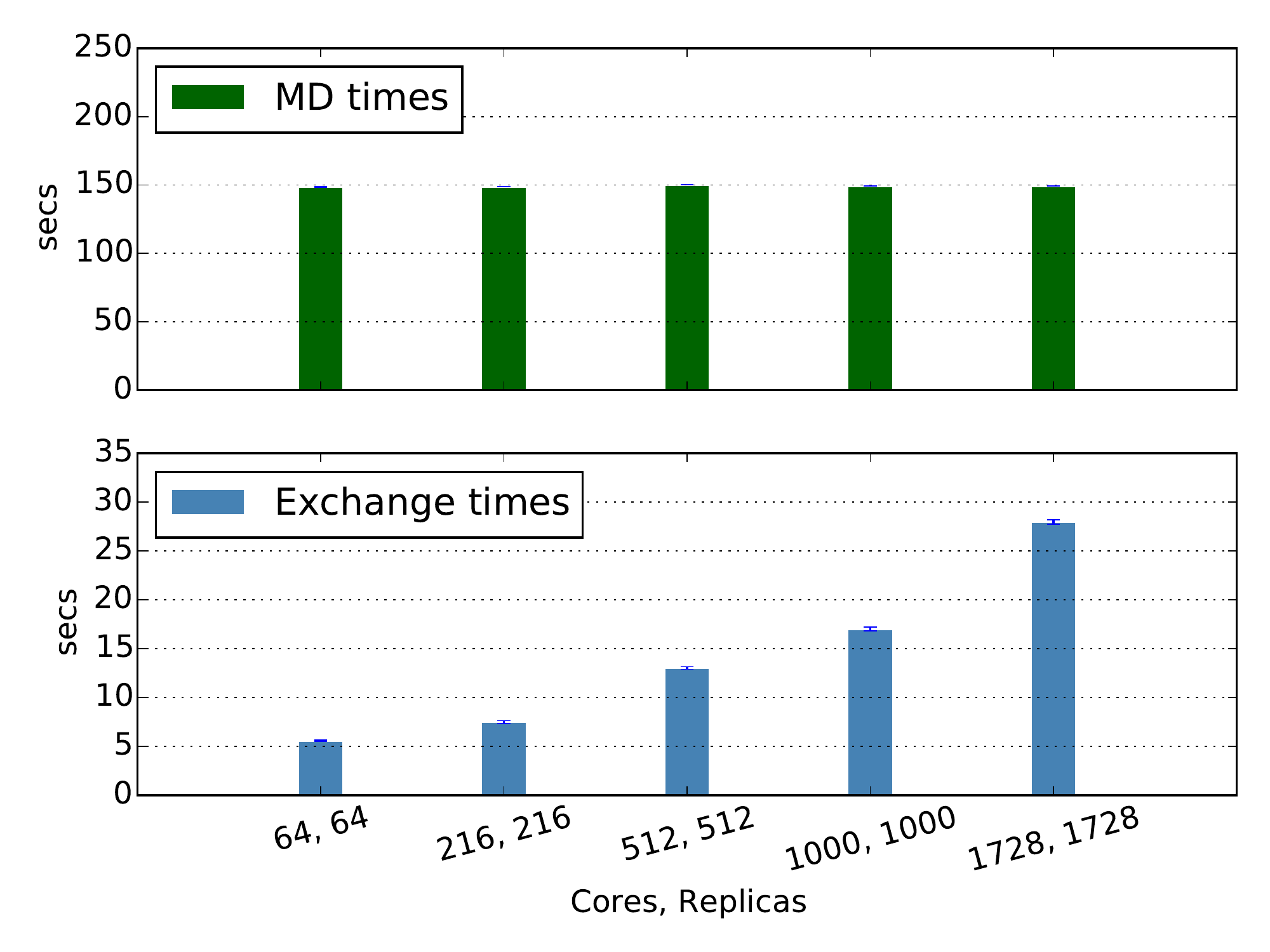}
\caption{\small{Experiments with NAMD engine. Decomposition of average simulation cycle times $T_{c}$ (in seconds) into MD simulation time and Exchange time for weak scaling scenario. Experiments are performed on SuperMIC supercomputer, using T-REMD. For MD simulation are used single-core replicas.}}
\label{fig:t-remd-namd}
\end{figure}

To demonstrate RepEx's ability to use different MD engines for REMD simulations we perform weak scaling experiments using T-REMD with NAMD engine. We run our experiments on SuperMIC, use NAMD-2.10 and perform a total of 4000 time-steps between exchanges. We perform runs with 64, 216, 512, 1000 and 1728 replicas. For each replica we use a single CPU core and we allocate enough cores to run all replicas concurrently (Execution Mode I). Results of these experiments are provided in Figure~\ref{fig:t-remd-namd}.

As expected, MD times for all cores/replicas pairs are nearly equal. Growth rate for exchange times can't be characterized as monomial.

\subsection{M-REMD performance characterization}

Similar to 1D-REMD experiments, we use alanine dipeptide to characterize M-REMD performance and 6000 simulation time-steps between exchanges. We perform weak and strong scaling experiments for TSU-REMD on Stampede supercomputer.

\begin{figure}[ht!]  \centering \includegraphics[width=0.45\textwidth]{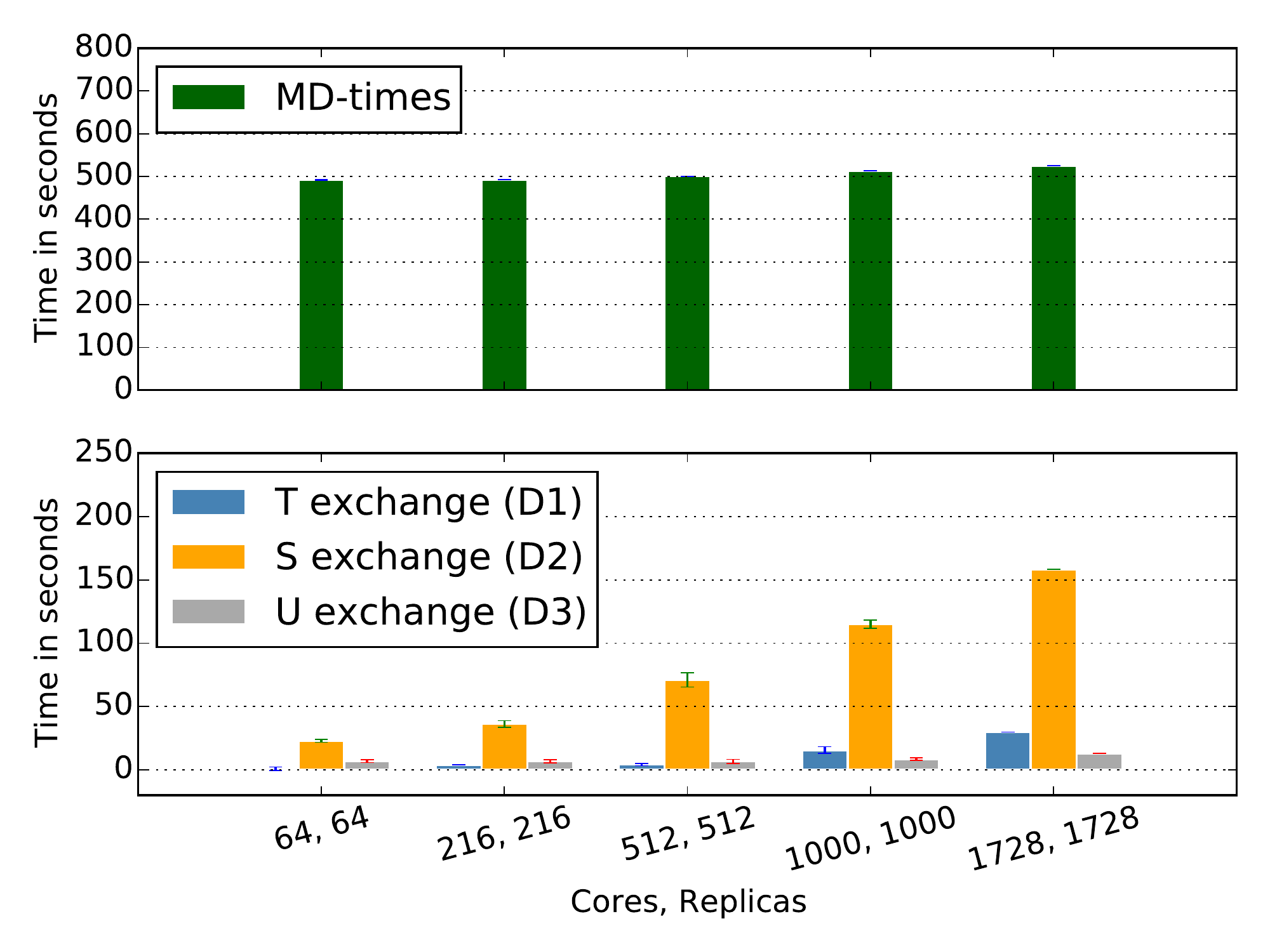} \caption{\small{Multi-dimensional REMD experiments with RepEx - weak scaling. TSU-REMD (Temperature, Salt concentration, Umbrella Sampling) on Stampede using Amber MD engine. For all simulation runs number of replicas is equal to the number of CPU cores and both vary from 64 to 1728. For all simulation runs are used single-core replicas. In figure is shown decomposition of average simulation cycle times $T_{c}$ (in seconds) into MD and exchange times.} }
\label{fig:tsu_weak} 
\end{figure}

{\bf Weak Scaling:} To characterize the weak scaling performance of M-REMD simulations, the number of replicas in each dimension is kept equal, thus as the number of replicas in one dimension varies from 4, 6, 8, 10 and 12, it results in the total number of replicas equal to 64, 216, 512, 1000 and 1728 respectively. We use Amber 12.0, and \textbf{sander} as Amber executable, as for each replica we use a single CPU core. The experiments are performed in Execution mode I, i.e., with enough cores to run all replicas concurrently.  Results of experiments are provided in Figure~\ref{fig:tsu_weak}.

For all simulation runs MD times are nearly identical: $\sim 495.0$ seconds. It is expected, since variation in the number of replicas should not affect MD time. 
 
We observe a nearly linear scaling for exchange timings in all three dimensions. While temperature and umbrella exchange timings are very similar, salt concentration exchange takes substantially more time. As mentioned in Subsection~\ref{exp.1d-remd}, for salt concentration exchange we use Amber to perform a single point energy calculations, which results in doubling of tasks and higher computational requirements for this exchange type.

\begin{figure}[ht!]
  \centering
  \subfigure[]{
  \includegraphics[width=0.22\textwidth]{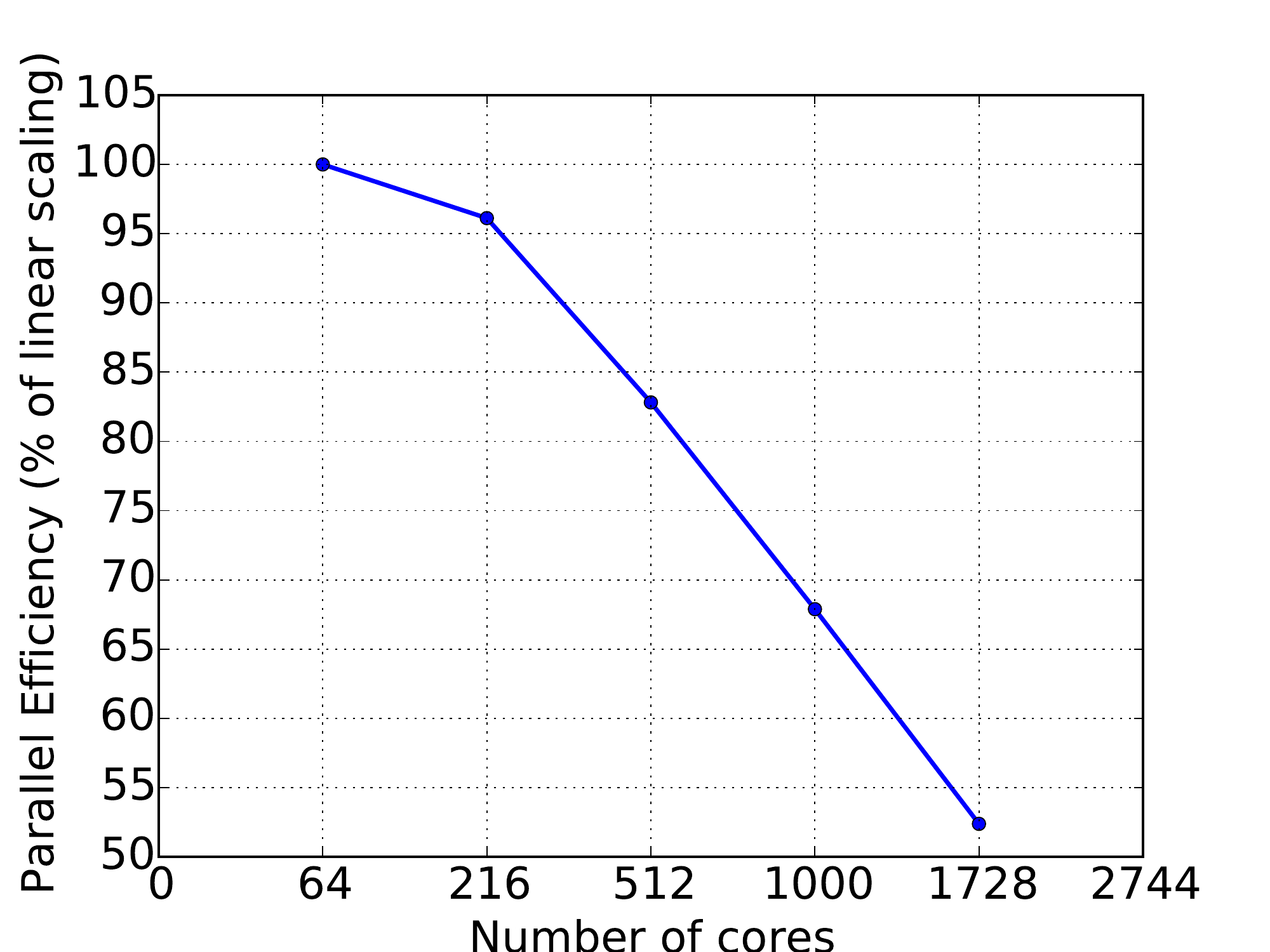}
  }
  \subfigure[]{
  \includegraphics[width=0.22\textwidth]{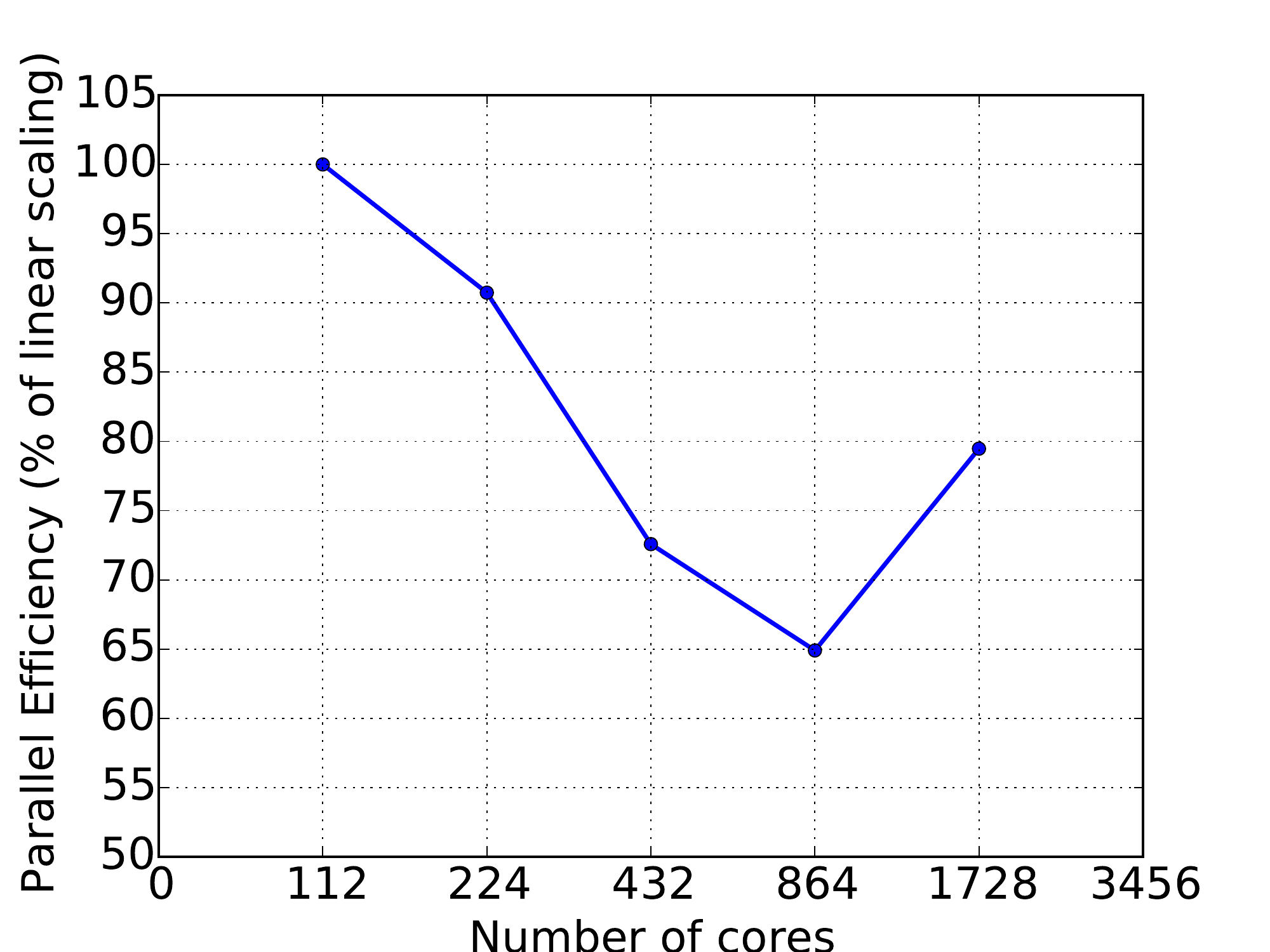}
  }
  \caption{\small{Parallel Efficiency (\% of linear scaling) for TSU-REMD on Stampede using Amber MD engine - (a) weak scaling, (b) strong scaling.} 
  }
  \label{fig:p_efficiency}
\end{figure}

Parallel Efficiency results are presented in Figure~\ref{fig:p_efficiency}(a). We observe rapid decrease in efficiency with increase in the number of cores. This can be explained by the influence of performance for salt concentration exchange. Despite that, for all core counts efficiency is above 50 \%.

\begin{figure}[ht!]
\centering
\includegraphics[width=0.45\textwidth]{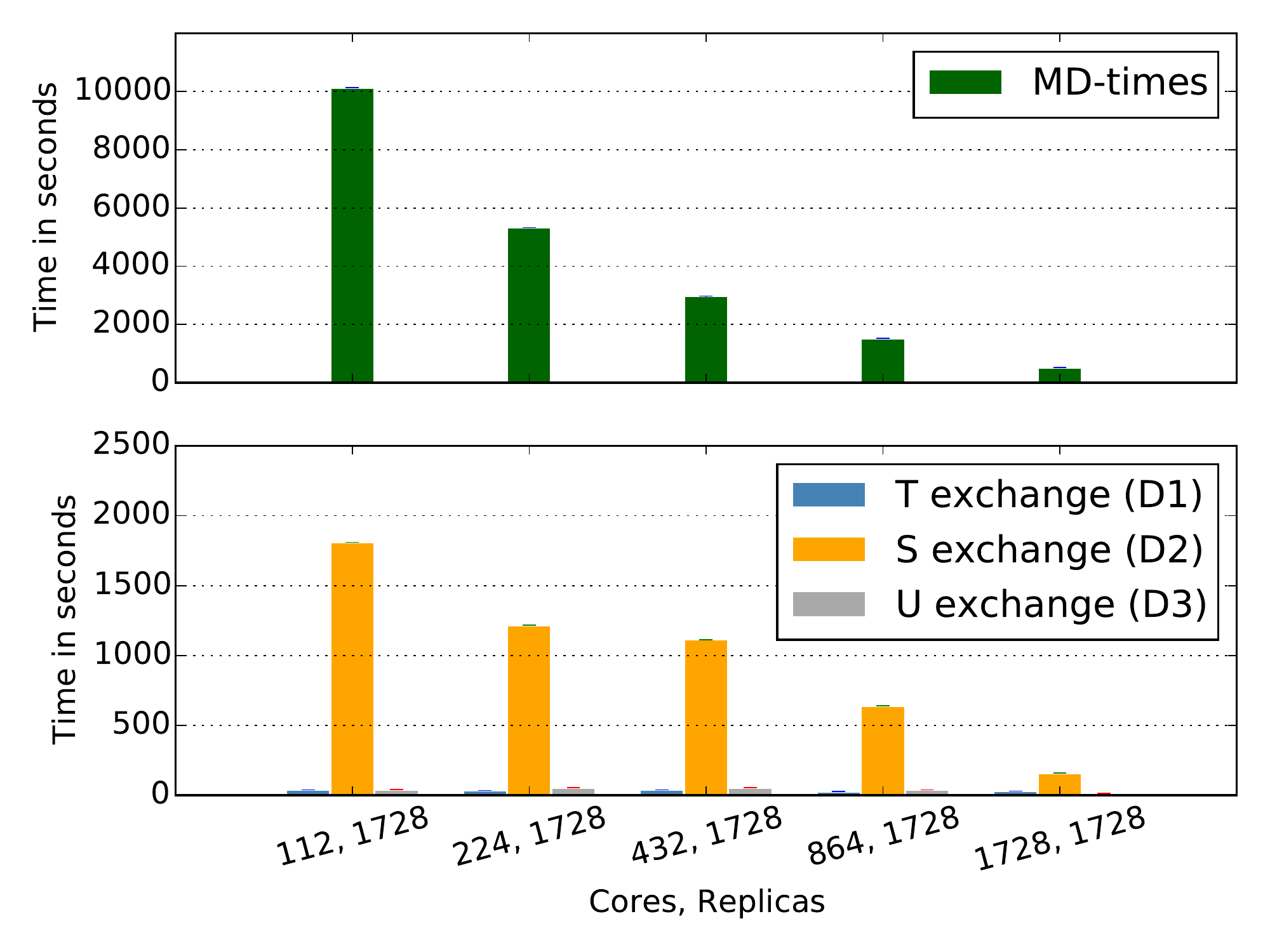}
\caption{\small{Multi-dimensional REMD experiments with RepEx: strong scaling. TSU-REMD (Temperature, Salt concentration, Umbrella Sampling) on Stampede using Amber MD engine. Number of replicas is fixed at 1728, but number of CPU cores is increased from 112 to 1728. For all runs are used single-core replicas. In figure are shown MD simulation and exchange times. RepEx enables users to vary the size of computational resources independently of the simulation size. Allocating more CPUs reduces the $T_{c}$.}    
}
\label{fig:tsu_strong}
\end{figure}

{\bf Strong Scaling:}  To characterize the strong scaling performance of M-REMD RepEx, the number of replicas is fixed at 1728 with 12 replicas in each dimension, but number of cores is varied: 112, 224, 432, 864 and 1728. Again, we use Amber 12.0, and \textbf{sander} as Amber executable, since for each replica a single CPU core is used.  The experiments are performed using Execution Mode II, as we have less cores than replicas for all cores/replicas pairs, except the last one. Results of these experiments are provided in Figure~\ref{fig:tsu_strong}.

As illustrated in Figure~\ref{fig:tsu_strong}, decrease in MD time is proportional to the number of cores: doubling of the number of CPU cores, results in decrease of MD time by nearly a half.

Exchange time in temperature exchange and umbrella exchange dimensions is almost equal for all numbers of CPU cores. This highlights the fact, that implementations of temperature exchange and umbrella exchange are very similar. Due to task launching delay and grouping of replicas by parameter values in each dimension, the exchanges largely overlap with MD. As a result, tasks which have finished simulation phase sooner can perform certain exchange procedures, before exchange is finalized. Compared to temperature exchange and umbrella exchange, salt concentration exchange times are significantly higher: at 112 cores, salt exchange time takes nearly 1800 seconds.

Parallel Efficiency results are presented in Figure~\ref{fig:p_efficiency}(b). As we can see, efficiency graph is non-linear. We observe decrease in efficiency up to the last data point where number of CPUs is equal to the number of replicas. For the last data point, efficiency increases. This behavior is caused by the MPI task scheduling issue of RP. In the next release of RepEx this issue will be addressed.   

\subsection{REMD with Multi-core Replicas} \label{exp.mcore.replicas}

To demonstrate RepEx capability to execute replicas using multiple cores and resulting reduction in total simulation time, we use solvated alanine dipeptide with 64366 atoms. We perform a total of 20000 time-steps between each exchange. Experiments are performed on Stampede using Amber 12.0 and pmemd.MPI as Amber executable for multi-core replicas and sander for single-core replicas. We use different executables, since pmemd.MPI can't be run on a single CPU core.

\begin{figure}[ht!]
\centering
\includegraphics[width=0.35\textwidth]{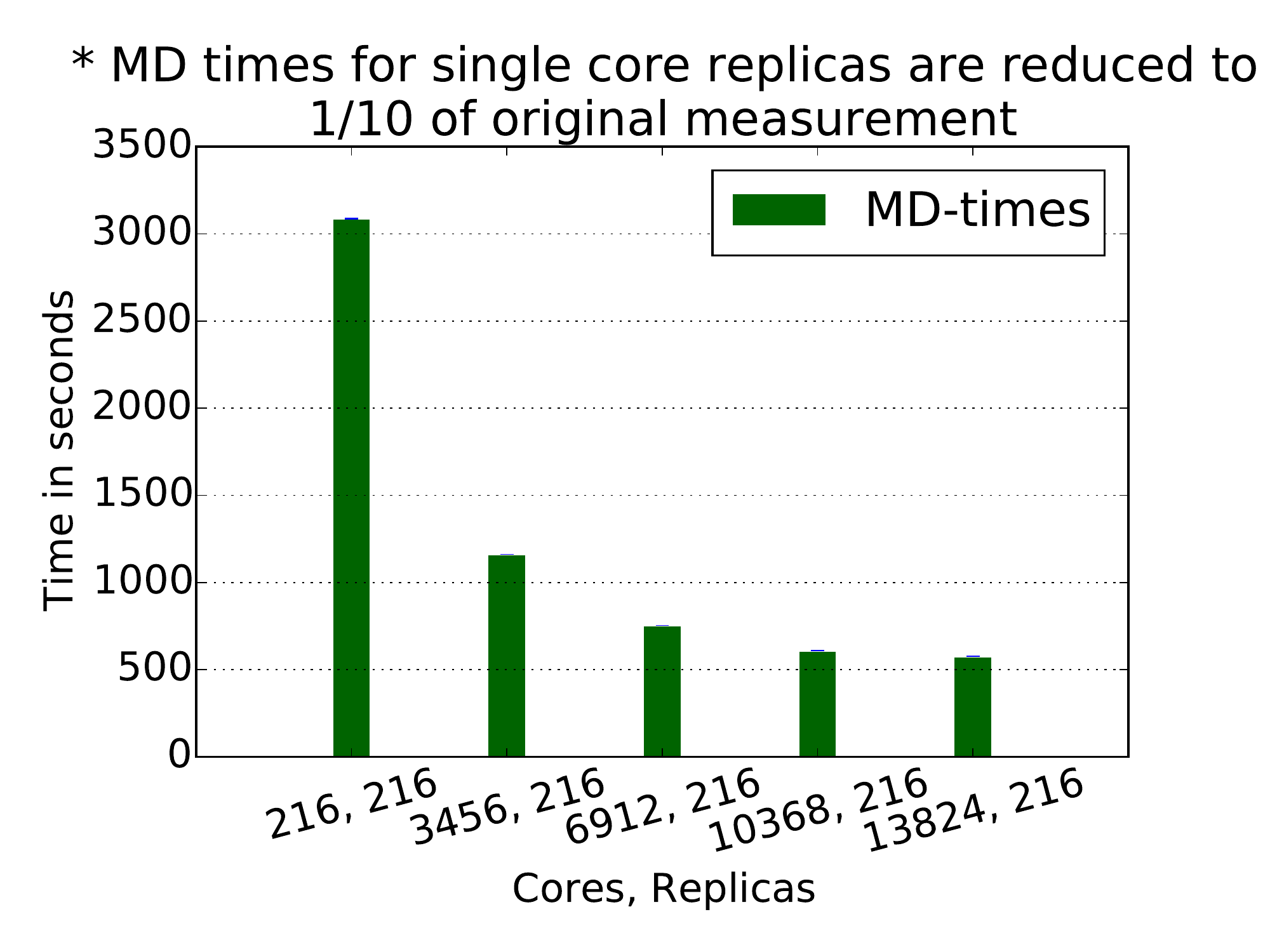}
\caption{\small{Multi-core replica experiments using TUU-REMD with Amber engine. MD times for weak scaling scenario. Experiments are performed on Stampede supercomputer. Number of replicas is fixed at 216, but number of CPUs per replicas is increased from 1 to 64.}}
\label{fig:multicore_3d}
\end{figure}

We perform weak scaling experiments using multi-core replicas and multi-dimensional TUU-REMD with one temperature dimension and two umbrella dimensions. We perform simulation runs with fixed number of replicas and change number of CPU cores per replica. For all runs we use 216 replicas, but the number of cores per replica varies from 1 to 64. Results of these experiments are provided in Figure~\ref{fig:multicore_3d}. 

We observe a substantial drop in MD times when we use multiple cores per replica. This is due to RepEx's ability to support replicas running over multi-core/multi-nodes, as well as using a highly efficient pmemd.MPI code. Further increase of CPU cores per replica doesn't demonstrate a linear behavior. This is not a limitation of the RepEx framework but attributable to the size of the alanine dipeptide, which although relatively larger than the earlier physical system, is small in absolute terms and thus makes it difficult to gain significant performance improvements by using more CPUs.

\subsection{Asynchronous REMD} \label{over}

In this subsection we compare utilization results of asynchronous RE pattern with synchronous RE pattern. For both patterns we use T-REMD with Amber engine and Execution Mode I. We use alanine dipeptide and perform 6000 time-steps for MD phase. We calculate utilization as:

\begin{equation} U = \frac{U_{pattern}}{U_{max}} \times 100 \% \end{equation} where:
\begin{compactitem}
  \item $U_{pattern}$ - utilization using (async/sync) RE pattern. Simulation time (ns/day) obtained per 1 CPU hour using this RE pattern. 
  \item $U_{max}$ - maximal (ideal) utilization, which is the amount of simulation time (ns/day) per 1 CPU hour, obtained assuming that CPU is used only to perform MD. 
\end{compactitem}

\begin{figure}[ht!]
  \centering
  \includegraphics[width=0.35\textwidth]{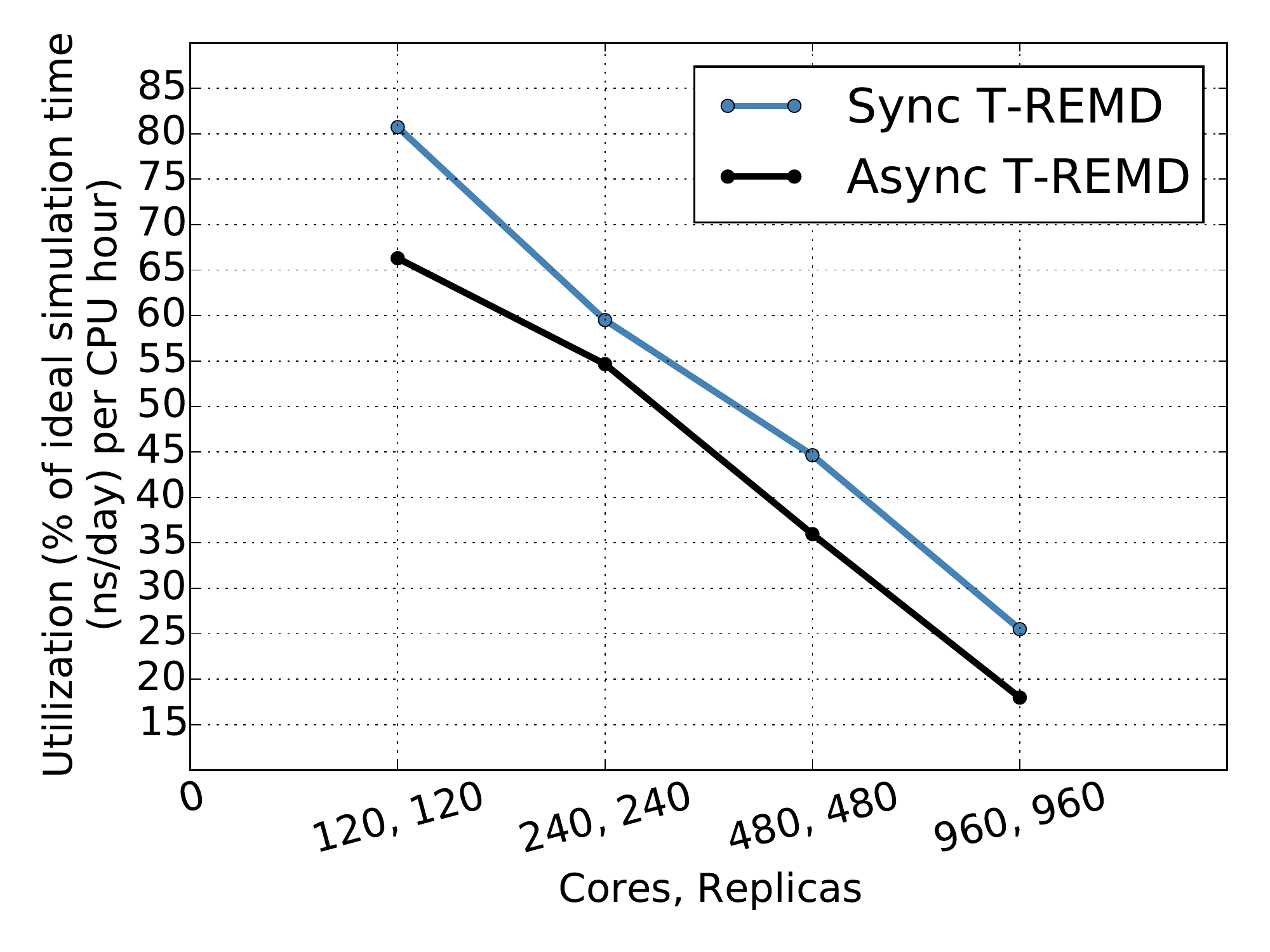}
  \caption{\small{Utilization for asynchronous and synchronous RE patterns using different replica counts. Utilization is a percentage of maximal (ideal) simulation time (ns/day) per CPU hour.  }
  }
  \label{fig:utilization}
\end{figure}

Figure~\ref{fig:utilization} shows that synchronous RE pattern results in higher utilization values invariant of the replica count.  There is approximately a 10\% difference between the two patterns, which though non-trivial is arguably an acceptable performance difference for the given advantages arising from asynchronous RE and resulting execution flexibility. It is worth mentioning, that for asynchronous RE, we use a fixed real time period as a criterion for transition of replicas to exchange phase. If other criterion, such as obtaining a certain energy value were to be used, we expect significantly better utilization results for asynchronous RE. We estimate that for large replica counts in Execution Mode II, the asynchronous RE pattern will out-perform synchronous RE pattern.

\section{Discussion and Conclusion} \label{discussion}

RepEx was designed to address functional, performance and usability requirements outlined in Section~\ref{requirements}. In section~\ref{experiments} we demonstrated capabilities of RepEx and characterized its performance for 1D and 3D REMD simulations. We saw the range of exchange parameters that it supports and the flexibility in their ordering (e.g., TUU versus TSU). Furthermore we saw RepEx supporting both Amber and NAMD with minimal conceptual or implementation changes. Last but not least, we saw the ability to utilize different RE patterns and different Execution Modes, thus providing a decoupling between exchange parameters (T/U/S), dimensionality and algorithm (sync. vs async.) on the one hand with resource management and execution details on the other. As such, it is accurate to say that RepEx satisfies the functional and usability requirements.

\begin{table*}[ht!]
  \begin{center}
  \begin{tabular}{| p{2.5cm} || p{1.2cm}| p{1.4cm} | p{1.6cm} | p{1.8cm} | p{1.8cm} | p{2.4cm} | p{1.6cm} |} 
  \hline
   & \textbf{Amber} & \textbf{Gromacs} & \textbf{LAMMPS} & \textbf{VCG async} & \textbf{CHARMM} & \textbf{Charm++ /NAMD MCA} & \textbf{RepEx}\\
  \hline
  \hline
  Max replicas & $\sim$2744 & $\sim$253 & 100 & 240 & 4096 & 2048 & 3584 \\ 
  \hline
  Max CPU cores & $\sim$5488 & $\sim$253 & 76800 & 1920 & 131072 & 524288 & 13824 \\ 
  \hline
  Fault tolerance & n/a & n/a & n/a & medium & n/a & n/a & medium \\  
  \hline
  MD engines & Amber & Gromacs & LAMMPS & IMPACT & CHARMM & NAMD & Amber, NAMD \\  
  \hline
  RE patterns & sync & sync & sync & sync, async & sync & sync & sync, async \\ 
  \hline
  Execution modes & low & low & low & medium & low & low & high \\ 
  \hline
  Nr. dims & 2 & 2 & 2 & 2 & 2 & 2 & 3 \\ 
  \hline
  Exchange params & 3 & 2 & 2 & 2 & 2 & 2 & 3 \\ 
  \hline
  \end{tabular}
  \end{center}
  \caption{\small{Comparison of molecular simulation software packages with integrated REMD capability. We characterize each of the seven packages based on eight features. For each feature we provide numerical value of that feature or one of three levels (low, medium, high). The only exception is "MD engines" feature, where we provide actual engine name.}}\label{table:compare}
\end{table*}

The two core design principles of RepEx are the separation of MD simulation engine from the implementation of RE algorithm, and the use of a pilot-based runtime which separates the algorithm and workload management from the resource management and runtime complexity.

As a consequence of the former, the integration of new MD simulation engines is significantly simplified and facilitates the reuse of RE patterns and Execution modes. We believe that this also lowers the barrier for development and testing of new REMD algorithms.  As a consequence of the second design principle, our implementation decouples execution specifics from the REMD algorithm and enables users to choose from the multiple execution options.  Collectively, this allowed us to introduce the concept of Replica Exchange pattern and demonstrate how RE patterns can be used interchangeably within RepEx. To the best of our knowledge none of the currently available REMD implementations have this capability.

The range of scalability (performance), generality (independent of MD engine) and flexibility (different configurations of exchanges) as demonstrated in the experiments validates the design of RepEx. 

In Table~\ref{table:compare} we have summarized the most important features of seven existing packages used for REMD, some of which are used by communities of hundreds, if not thousands of users. In this table we have included three popular MD simulation engines, namely Amber, LAMMPS and Gromacs that have been extended to provide RE capabilities, and four REMD packages that have been designed to be external to MD engines. Some of these were reviewed in Section~\ref{remd.landscape}.

As can be seen from Table~\ref{table:compare}, a majority of the packages are designed to address a subset of features we identified as necessary in order to be flexible and general purpose. Many packages have eschewed generality for performance. For example, Charm++/NAMD MCA package can utilize O(100,000) cores but does not provide flexible resource utilization nor asynchronous exchange capabilities. On the other hand, VCG RE package is one of the few packages, which supports asynchronous RE but it has limited scalability (both in the number of replicas and cores that it has been used for) and is tightly coupled to IMPACT which is not an open source MD engine. Similar to most other existing solutions, both VCG and Charm++/NAMD are limited in the number of exchange parameters as well as in flexibility in ordering of exchange parameters.

Clearly a balance between performance and functional requirements needs to be maintained.  On the evidence of Table~\ref{table:compare} we believe that RepEx provides an optimal balance.  As evidenced by the careful requirements analysis, design and implementation considerations, RepEx embodies the sound systems engineering principles along with software engineering practices.  RepEx is now being used for algorithmically innovative molecular science simulations~\cite{ct500776j}.
    
Our preliminary results show that RepEx can easily be extended to support use of GPUs for simulation phase. Based upon significant end-user request, support for GPUs is already available on Stampede and will be extended to other machines, such as Blue Waters~\cite{blue.waters} and Titan~\cite{titan}.

There are some obvious extensions to the current RepEx framework: First, single point energy calculations for salt concentration exchange can be implemented. Next, a number of additional exchange parameters can be added to support other types of multi-dimensional REMD simulations (for example pH exchange). Third, support for additional MD simulation engines might be introduced. Finally, RepEx can be extended to use multiple HPC resources simultaneously for a single REMD simulation.

\bibliographystyle{ieeetr}

\section*{Acknowledgment}

This work is supported by NSF CHE-1265788. We acknowledge allocation TG-MCB090174 for computing time on XSEDE allocated resources. We acknowledge NSF ACI 1516469 and ACI 1515572 (Cheatham) for time on Blue Waters machine.

\bibliography{references,radical_publications}

\begin{thebibliography}{10}

\bibitem{swendsen1986replica}
R.~H. Swendsen and J.-S. Wang, ``Replica monte carlo simulation of
  spin-glasses,'' {\em Physical Review Letters}, vol.~57, no.~21, p.~2607,
  1986.

\bibitem{sugita1999replica}
Y.~Sugita and Y.~Okamoto, ``Replica-exchange molecular dynamics method for
  protein folding,'' {\em Chemical physics letters}, vol.~314, no.~1,
  pp.~141--151, 1999.

\bibitem{fukunishi2002hamiltonian}
H.~Fukunishi, O.~Watanabe, and S.~Takada, ``On the hamiltonian replica exchange
  method for efficient sampling of biomolecular systems: application to protein
  structure prediction,'' {\em The Journal of chemical physics}, vol.~116,
  no.~20, pp.~9058--9067, 2002.

\bibitem{meng2010constant}
Y.~Meng and A.~E. Roitberg, ``Constant ph replica exchange molecular dynamics
  in biomolecules using a discrete protonation model,'' {\em Journal of
  chemical theory and computation}, vol.~6, no.~4, pp.~1401--1412, 2010.

\bibitem{Amber}
R.~Salomon-Ferrer, D.~A. Case, and R.~C. Walker, ``An overview of the amber
  biomolecular simulation package,'' {\em Wiley Interdisciplinary Reviews:
  Computational Molecular Science}, vol.~3, no.~2, pp.~198--210, 2013.

\bibitem{namd2005}
J.~C. Phillips, R.~Braun, W.~Wang, J.~Gumbart, E.~Tajkhorshid, E.~Villa,
  C.~Chipot, R.~D. Skeel, L.~Kale, and K.~Schulten, ``Scalable molecular
  dynamics with namd,'' {\em Journal of computational chemistry}, vol.~26,
  no.~16, pp.~1781--1802, 2005.

\bibitem{gromacs1995}
H.~J. Berendsen, D.~van~der Spoel, and R.~van Drunen, ``Gromacs: A
  message-passing parallel molecular dynamics implementation,'' {\em Computer
  Physics Communications}, vol.~91, no.~1, pp.~43--56, 1995.

\bibitem{repex-code}
``Repex on github.'' \url{https://github.com/radical-cybertools/radical.repex}.
\newblock Accessed: 2015-11-11.

\bibitem{review_radicalpilot_2015}
A.~Merzky, M.~Santcroos, M.~Turilli, and S.~Jha, ``{RADICAL-Pilot: Scalable
  Execution of Heterogeneous and Dynamic Workloads on Supercomputers},'' 2015.
\newblock (under review) \url{http://arxiv.org/abs/1512.08194}.

\bibitem{Jiang_JChemTheoryComput_2012_v8_p4672}
W.~Jiang, Y.~Luo, L.~Maragliano, and B.~Roux, ``{Calculation of free energy
  landscape in multi-dimensions with Hamiltonian-exchange umbrella sampling on
  petascale supercomputer},'' {\em J. Chem. Theory Comput.}, vol.~8,
  pp.~4672--4680, 2012.

\bibitem{charmm}
B.~R. Brooks, C.~L. Brooks, A.~D. MacKerell, L.~Nilsson, R.~J. Petrella,
  B.~Roux, Y.~Won, G.~Archontis, C.~Bartels, S.~Boresch, {\em et~al.},
  ``Charmm: the biomolecular simulation program,'' {\em Journal of
  computational chemistry}, vol.~30, no.~10, pp.~1545--1614, 2009.

\bibitem{jiang2014}
W.~Jiang, J.~C. Phillips, L.~Huang, M.~Fajer, Y.~Meng, J.~C. Gumbart, Y.~Luo,
  K.~Schulten, and B.~Roux, ``Generalized scalable multiple copy algorithms for
  molecular dynamics simulations in namd,'' {\em Computer physics
  communications}, vol.~185, no.~3, pp.~908--916, 2014.

\bibitem{shalongo}
W.~Shalongo, L.~Dugad, and E.~Stellwagen, ``Distribution of helicity within the
  model peptide acetyl (aaqaa) 3amide,'' {\em Journal of the American Chemical
  Society}, vol.~116, no.~18, pp.~8288--8293, 1994.

\bibitem{2013-xsede-cdi}
B.~K. Radak, M.~Romanus, E.~Gallicchio, T.-S. Lee, O.~Weidner, N.-J. Deng,
  P.~He, W.~Dai, D.~M. York, R.~M. Levy, and S.~Jha, ``{A Framework for
  Flexible and Scalable Replica-Exchange on Production Distributed CI},'' XSEDE
  '13, pp.~26:1--26:8, 2013.

\bibitem{ct500776j}
B.~K. Radak, M.~Romanus, T.-S. Lee, H.~Chen, M.~Huang, A.~Treikalis,
  V.~Balasubramanian, S.~Jha, and D.~M. York, ``{Characterization of the
  Three-Dimensional Free Energy Manifold for the Uracil Ribonucleoside from
  Asynchronous Replica Exchange Simulations},'' {\em Journal of Chemical Theory
  and Computation}, vol.~11, no.~2, pp.~373--377, 2015.
\newblock \url{http://dx.doi.org/10.1021/ct500776j}.

\bibitem{impact2005}
J.~L. Banks, H.~S. Beard, Y.~Cao, A.~E. Cho, W.~Damm, R.~Farid, A.~K. Felts,
  T.~A. Halgren, D.~T. Mainz, J.~R. Maple, {\em et~al.}, ``Integrated modeling
  program, applied chemical theory (impact),'' {\em Journal of computational
  chemistry}, vol.~26, no.~16, pp.~1752--1780, 2005.

\bibitem{Xia_JComputChem_2015_v36_p1772}
J.~Xia, W.~F. Flynn, E.~Gallicchio, B.~W. Zhang, P.~He, Z.~Tan, and R.~M. Levy,
  ``{Large-scale asynchronous and distributed multidimensional replica exchange
  molecular simulations and efficiency analysis},'' {\em J. Comput. Chem.},
  vol.~36, pp.~1772--1785, 2015.

\bibitem{Bergonzo_JChemTheoryComput_2014_v10_p492}
C.~Bergonzo, N.~M. Henriksen, D.~R. Roe, J.~M. Swails, A.~E. Roitberg, and
  T.~E. Cheatham~III, ``{Multidimensional replica exchange molecular dynamics
  yields a converged ensemble of an RNA tetranucleotide},'' {\em J. Chem.
  Theory Comput.}, vol.~10, pp.~492--499, 2014.

\bibitem{Panteva_BookChap_MultiscaleRNAEnzym_2015_v553_p335}
M.~T. Panteva, T.~Dissanayake, H.~Chen, B.~K. Radak, E.~R. Kuechler, G.~M.
  Giamba\c{s}u, T.-S. Lee, and D.~M. York, {\em Multiscale Methods for
  Computational RNA Enzymology}, ch.~14.
\newblock Elsevier, 2015.

\bibitem{Ensing_AccChemRes_2006_v39_p73}
B.~Ensing, M.~De~Vivo, Z.~Liu, P.~Moore, and M.~L. Klein, ``{Metadynamics as a
  tool for exploring free energy landscapes of chemical reactions},'' {\em Acc.
  Chem. Res.}, vol.~39, no.~2, pp.~73--81, 2006.

\bibitem{Vanden-Eijnden_JComputChem_2009_v30_p1737}
E.~Vanden-Eijnden, ``{Some recent techniques for free energy calculations},''
  {\em J. Comput. Chem.}, vol.~30, no.~11, pp.~1737--1747, 2009.

\bibitem{Dissanayake_Biochemistry_2015_v54_p1307}
T.~Dissanayake, J.~M. Swails, M.~E. Harris, A.~E. Roitberg, and D.~M. York,
  ``{Interpretation of pH-Activity Profiles for Acid-Base Catalysis from
  Molecular Simulations},'' {\em Biochemistry}, vol.~54, pp.~1307--1313, 2015.

\bibitem{swadling2015structure}
J.~B. Swadling, D.~W. Wright, J.~L. Suter, and P.~V. Coveney, ``Structure,
  dynamics, and function of the hammerhead ribozyme in bulk water and at a clay
  mineral surface from replica exchange molecular dynamics,'' {\em Langmuir},
  vol.~31, no.~8, pp.~2493--2501, 2015.

\bibitem{review_pilotreview_2013}
M.~Turilli, M.~Santcroos, and S.~Jha, ``{A Comprehensive Perspective on
  Pilot-Jobs},'' 2015.
\newblock \url{http://arxiv.org/abs/1508.04180}.

\bibitem{xsede}
``Extreme science and engineering discovery environment.''
  \url{https://www.xsede.org/resources/overview}.
\newblock Accessed: 2015-11-11.

\bibitem{Lee_JChemTheoryComput_2013_v9_p153}
T.-S. Lee, B.~K. Radak, A.~Pabis, and D.~M. York, ``{A new maximum likelihood
  approach for free energy profile construction from molecular simulations},''
  {\em J. Chem. Theory Comput.}, vol.~9, pp.~153--164, 2013.

\bibitem{Lee_JChemTheoryComput_2014_v10_p24}
T.-S. Lee, B.~K. Radak, M.~Huang, K.-Y. Wong, and D.~M. York, ``{Roadmaps
  through free energy landscapes calculated using the multidimensional vFEP
  approach},'' {\em J. Chem. Theory Comput.}, vol.~10, pp.~24--34, 2014.

\bibitem{repex-experiments}
``Repex experiments.''
  \url{https://github.com/radical-cybertools/radical.repex/blob/master/EXPERIMENTS.md}.
\newblock Accessed: 2015-11-11.

\bibitem{blue.waters}
``Blue waters supercomputer at the national center for supercomputing
  applications.'' \url{https://bluewaters.ncsa.illinois.edu/blue-waters}.
\newblock Accessed: 2015-11-11.

\bibitem{titan}
``Titan supercomputer - oak ridge leadership computing facility.''
  \url{https://www.olcf.ornl.gov/titan/}.
\newblock Accessed: 2015-11-11.

\end{thebibliography}

\end{document}